\documentclass[natbib]{svjour3}                     
\smartqed  
\input{epsf}
\usepackage{epsfig}
\usepackage{graphicx}
\usepackage{graphics}
\usepackage{amsmath}

 \usepackage{mathptmx}      
 \usepackage{aps-bibstyle}  
\def\be{\begin{equation}}
\def\ee{\end{equation}}
\def\bdm{\begin{displaymath}}
\def\edm{\end{displaymath}}

\begin{document}

\title{Kappa distributions: theory and applications in space plasmas
}

\titlerunning{Kappa distributions in space plasmas}        

\author{V. Pierrard$^{1,2}$         \and
        M. Lazar$^{3,4}$ 
}

\authorrunning{V. Pierrard and M. Lazar} 

\institute{V. Pierrard \at
              $^1$ Belgian Institute for Space Aeronomy, av. circulaire 3,
              B-1180 Brussels, Belgium \\
              $^2$ Center for Space Radiations, Universit\'e Catholique de
              Louvain, chemin du cyclotron 2, B-1348 Louvain-La-Neuve,
              Belgium\\
              \email{viviane.pierrard@oma.be}           
           \and
           M. Lazar \at
              $^3$ Research Department - Plasmas with Complex
                Interactions, \\Ruhr-Universit\"at Bochum, D-44780 Bochum\\
              \email{mlazar@tp4.rub.de}\\
              $^4$ Center for Plasma Astrophysics, Celestijnenlaan 200B,
              3001 Leuven, Belgium
}

\date{Received: date / Accepted: date}

\maketitle

\begin{abstract}
Particle velocity distribution functions (VDF) in space plasmas
often show non Maxwellian suprathermal tails decreasing as a power
law of the velocity. Such distributions are well fitted by the
so-called Kappa distribution. The presence of such distributions in
different space plasmas suggests a universal mechanism for the
creation of such suprathermal tails. Different theories have been
proposed and are recalled in this review paper. The suprathermal
particles have important consequences concerning the acceleration
and the temperature that are well evidenced by the kinetic approach
where no closure requires the distributions to be nearly
Maxwellians. Moreover, the presence of the suprathermal particles
take an important role in the wave-particle interactions.
\keywords{Kappa distributions \and space plasmas \and kinetic models \and waves and instabilities}
\end{abstract}

\section{Introduction}
\label{intro} Nonthermal particle distributions are ubiquitous at
high altitudes in the solar wind and many space plasmas, their
presence being widely confirmed by spacecraft measurements
\citep{mo68, fe75, pi87, Maks97a, zoug08}. Such deviations from the
Maxwellian distributions are also expected to exist in any
low-density plasma in the Universe, where binary collisions of
charges are sufficiently rare. The suprathermal populations are well
described by the so-called Kappa ($\kappa$-) or generalized
Lorentzian velocity distributions functions (VDFs), as shown for the
first time by \citet{v68}. Such distributions have high energy tails
deviated from a Maxwellian and decreasing as a power law in particle
speed:

\begin{equation}
f_i^{\kappa}(r,v)=\frac{n_{i}}{2\pi(\kappa w_{\kappa i}^2)^{3/2}}
\frac{\Gamma(\kappa+1)}{\Gamma(\kappa-1/2) \Gamma(3/2)} \left(
1+\frac{v^2}{\kappa w_{\kappa i}^2} \right)^{-(\kappa+1)} \label{e1}
\end{equation}
where $w_{\kappa i}=\sqrt{(2\kappa-3)kT_{i}/\kappa m_i}$ is the
thermal velocity, $m_i$ the mass of the particles of species $i$,
$n$ their number density, $T$ their equivalent temperature, $v$ the
velocity of the particles, and $\Gamma(x)$ is the Gamma function.
The spectral index must take sufficiently large values $\kappa >
3/2$ to keep away from the critical value $\kappa_c = 3/2$, where
the distribution function (\ref{e1}) collapses and the equivalent
temperature is not defined. The value of the index $\kappa$ determines the slope of the energy
spectrum of the suprathermal particles forming the tail of the VDF, as illustrated on Fig. 1.
In the limit $\kappa \rightarrow\infty$, the Kappa function
degenerates into a Maxwellian. Note also that different mathematical
definitions of Kappa distributions are commonly used and various
authors characterize the power law nature of suprathermal tails in
different ways.

Here some comments are necessary because the conventional Kappa
distribution function (\ref{e1}) used to fit and describe high
energy tails incorporates macroscopic parameters defined by the
lowest moments of the distribution function. We follow some suggestive
explanations from \cite{hel09}, and first remember that $w_{\kappa
i}$ in (\ref{e1}) was originally stated by \cite{v68} to be the most
probable particle speed. Thus, a characteristic (nonrelativistic) kinetic
energy, $W = mw_{\kappa i}^2 /2$, can be associated to the most
probable speed, and by considering the second moment of the
distribution function $U=\int d{\bf v} f_i^{\kappa}mv^2/2$, the mean
energy per particle reads $W_m \equiv U/N = W(3 \kappa/2)/ (\kappa
-3/2)$. Later, \citet{Form73} have introduced a plasma temperature
related to the mean energy per particle $k_B T = (2/3)W_m = m
w_{\kappa i}^2 \kappa /(2 \kappa -3)$, which is exactly the
equivalent temperature proposed by \citet{l82} and \citet{cha91} by
relating to the average energy $k_BT = m$$<$$v^2$$>$$/3$. On this
basis, it was also shown that, in a Kappa distributed plasma, the
Debye length is less than in a Maxwellian plasma, $\lambda_{\kappa}
= \lambda (2\kappa -3)/(2 \kappa -1)$. Such a temperature
definition, making use of equipartition of energy, although
appropriate for an equilibrium Maxwellian distribution, is not
strictly valid for a Kappa distribution, but there are practical
advantages to using such an equivalent kinetic temperature, which
can be a useful concept already accepted in practice for
non-Maxwellian distributions (see \cite{hel09} and references
therein).

Furthermore, generalizations of thermodynamics based on the Tsallis
nonextensive entropy formalism \citep{Tsal95} have been used for
several decades (\citet{Liva09} and references therein). The
family of kappa distributions result from a new generalized
Lorentzian statistical mechanics formulated for a collisionless
plasmas far from thermal (Boltzmann-Maxwell) equilibrium but
containing fully developed turbulence in quasistationary equilibrium
\citep{treu99a, treu99b, Leub02, fg06, gf06, treu08}. Thus, in phase
space, the power-law distributions of Kappa type describe marginally
stable Gibbsian equilibria, and the parameter $\kappa$ controls the
strength of the plasma particle correlation in the turbulent field
fluctuations \citep{Hase85, treu08}. In such systems the temperature
is redefined on the basis of a superadditive (superextensive)
entropy because the interdependence of subsystems contributes an
extra amount to entropy \citep{treu08}.

Such Kappa functions give the best fit to the observed velocity
distribution functions, using only 3 parameters (the number density
$n$, the temperature $T$ and the parameter $\kappa$ characterizing
the suprathermal tails). A sum of two Maxwellians can also represent
distributions with enhanced suprathermal tails, but they need 4
parameters ($n_1$, $T_1$ and $n_2$, $T_2$ representing the number
density and temperatures of the two populations) and they generally
give less good fits than Kappa functions \citep{zoug04}.

\begin{figure}
\begin{center}
\includegraphics[height=8cm]{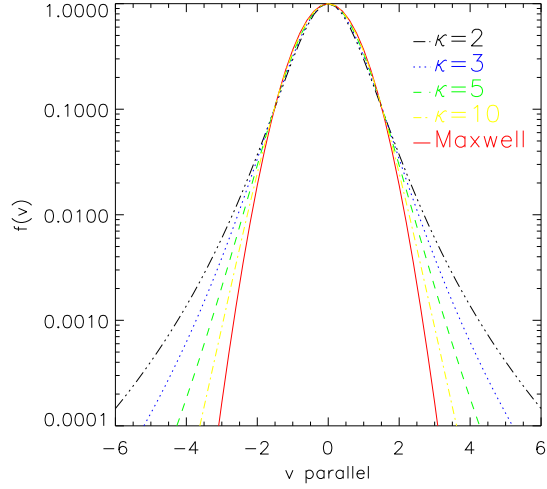}
\end{center}
\caption{The Kappa velocity distribution function for different
values of the kappa parameter.} \label{fig2.1}
\end{figure}

Considering the suprathermal particles has important consequences for space plasmas. For instance, an isotropic Kappa distribution (instead of a
Maxwellian) in a planetary or stellar exosphere leads to a number
density $n(r)$ decreasing as a power law (instead of exponentially)
with the radial distance $r$ and a temperature $T$ increasing with
the radial distance (instead of being constant), as shown by the expressions given in
Table 1. $R(r)$ is the potential energy containing the effects of
the gravitation, the electrostatic and centrifugal force. $v_e$ is the
escape velocity and $A_k$ is the fraction of Gamma functions
appearing in the Kappa VDF.
Considering particles escaping as planetary or stellar wind, the Kappa distribution yield higher flux
than a Maxwellian, since more suprathermal particles are able to escape.

\begin{table}
\caption{Comparison of different analytical expressions for a
Maxwellian and a Kappa VDF.}
\begin{center}
\begin{tabular}{llll} \hline\noalign{\smallskip}
Parameter & Maxwellian & Kappa\\
\noalign{\smallskip}\hline\noalign{\smallskip}
Number density & $n(r)=n_0 \exp\left(-\frac{R(r)}{w^2}\right)$ & $n(r)=n_0(r)\left( 1+\frac{R(r)}{\kappa w^2}\right)^{-\kappa+1/2}$ \\
Temperature& $T(r)=T_0$ & $T(r)=T_0\frac{\kappa}{\kappa-3/2}\left( 1+\frac{R(r)}{\kappa w^2}\right) $ \\
Escaping flux& $F(r)=\frac{n_0 w (1+v_e^2/w^2)}{2 \pi^{1/2}}
\exp\left(-\frac{v_e^2}{w^2}\right)$&
$F(r)=\frac{n_0  A_k w (1+v_e^2/w^2)}{4(\kappa-1)\kappa^{1/2}[1+v_e^2/(\kappa w^2)]^{\kappa}}$\\
\noalign{\smallskip}\hline
\end{tabular}
\end{center}
\end{table}

This review is organized in the following fashion. Reports of
measurements or indirect detections of Kappa distributions in our
interplanetary space are reviewed in the next section. In Section 3,
we identify the mechanisms made responsible for the occurrence of
nonthermal populations in different environments. Representative
theories and scenarios developed for Kappa distributed plasmas are
discussed in Section 4. In Section 5 we make a short overview of the
dynamics and dispersion properties of Kappa distributions including
the recent results on the stability of anisotropic plasmas and
kinetic instabilities. The impact and favorable perspectives for
these distributions are discussed in Section 6.

\section{Detection of Kappa distributions}

Distributions with suprathermal tails have been observed in various
space plasmas. Kappa distributions with $2< \kappa < 6$ have been
found to fit the observations and satellite data  in the solar wind
\citep{Gloe92, Maks97a}, the terrestrial magnetosphere
\citep{Gloe87}, the terrestrial plasmasheet \citep{Bame67, Chri88,
Chri89, Klet03}, the magnetosheath \citep{Form73}, the radiation
belts \citep{PiLe96b, xiao08}, the magnetosphere of other planets
like Mercury \citep{Chri87}, the plasmasheet of Jupiter
\citep{Colal95}, the magnetosphere of Jupiter \citep{Krim81, Mauk04}, Saturn
\citep{Krim83, Schi08, Dial09}, Uranus \citep{Krim86}, Neptune \citep{Mauk91}, and
even on Titan \citep{Haye07} and in the Io plasma torus as observed
by Ulysses \citep{Meye95}, Cassini \citep{Stef04} and the Hubble
Space Telescope \citep{Reth03}.

\begin{figure}
\begin{center}
\includegraphics[width=5.cm]{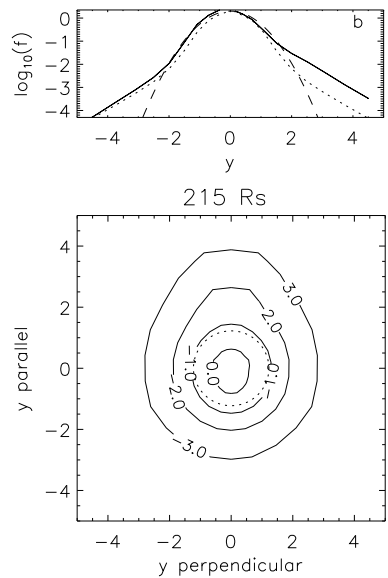}
\end{center}
\caption{Velocity distribution function observed by the satellite
WIND at 1 UA (215 Rs) for high speed solar wind electrons. Bottom
panel: Isocontours in the plane of velocities (normalized to the
thermal velocity) parallel and perpendicluar to the interplanetary
magnetic field. Top panel: Parallel (solid line) and perpendicular
(dotted line) cross section of the observed VDF. The dashed line
represents the Maxwellian distribution that well fits the core of
observed VDF. \citep{pml99}} \label{windobs}
\end{figure}

In the solar wind, electron velocity distributions are characterized
by a thermal core and a halo suprathermal population
\citep{Pier01b}, as illustrated on Fig. \ref{windobs}. These electron VDF are also characterized by a strahl component
aligned with the interplanetary magnetic field.
Electron VDF
measured by Ulysses have been fitted with Kappa functions by
\citet{Maks97a}. They show a global anticorrelation between the
solar wind bulk speed and the value of the parameter $\kappa$, that
supports the kinetic theoretical result that the suprathermal
electrons influence the solar wind acceleration \citep{Maks97b}.
Solar wind particle VDF observed by CLUSTER have also been fitted by
the generalized Kappa function \citep{qure03}. Radial evolution of
nonthermal electron populations in the low-latitude solar wind with
Helios, Cluster, and Ulysses observations shows that the relative
number of strahl electrons is decreasing with radial distance,
whereas the relative number of halo electrons is increasing
\citep{Stve09}. Observations of electron suprathermal tails in the
solar wind suggest their existence in the solar corona, since the
electron mean free path in the solar wind is around 1 AU. The ion charge measurements stated by Ulysses were found to be
consistent with coronal Kappa VDF of electrons with kappa index
ranging between 5 and 10 \citep{Koal96}.

To be able to measure the suprathermal electron parameters in space
plasmas, the quasi-thermal noise spectroscopy was implemented with
Kappa distributions using in situ Ulysses/\ URAP radio measurements
in the solar wind \citep{zoug08}. This noise is produced by the
quasi-thermal fluctuations of the electrons and by the Doppler-shifted
thermal fluctuations of the ions. A sum of two Maxwellians has
extensively been used but the observations have shown that the electrons
are better fitted by a kappa distribution function \citep{lechat09}.

 Solar wind ion ($^{20}$Ne,
$^{16}$O and $^4$He) distribution functions measured by WIND and
averaged over several days have also been fitted by Kappa functions
\citep{Coll96}. Low values of $\kappa$ (between 2.4 and 4.7) are
obtained due to the presence of high suprathermal tails. Suprathermal solar wind particles were also measured in H$^+$, He$^{++}$, and He$^+$ distribution functions during corotating interaction region (CIR) events observed by WIND at 1 AU \citep{Chot00}.

\section{Generation of Kappa distributions in space plasmas}

Various mechanisms have been proposed to explain the origin of the
suprathermal tails of the VDFs and occurrence of Kappa-like
distributions in the solar wind, the corona and other space plasmas.
The first one was suggested by \citet{Hase85} who showed that a
plasma immersed in a suprathermal radiation field suffers
velocity-space diffusion which is enhanced by the photon-induced
Coulomb-field fluctuations. This enhanced diffusion universally
produces a power-law distribution.

\citet{Coll93} uses random walk jumps in velocity space whose
path lengths are governed by a power or L\'evy flight probability
distribution to generate Kappa-like distribution functions. The
adiabatic transport of suprathermal distributions modeled by Kappa
functions is studied in \citet{Coll95}. The same author shows that
space plasmas are dynamic systems where the energy is not fixed, so
that the maximum entropy should not be considered \citep{Coll04}.

\citet{Treu01} developed a kinetic theory to show that Kappa-like
VDFs correspond to a particular thermodynamic equilibrium state
\citep{treu99b, treu04}. A new kinetic theory Boltzmann-like
collision term including correlations was proposed. In equilibrium (turbulent but stable state far from thermal equilibrium),
it yields the one-particle distribution function in the form of a
generalized Lorentzian resembling but not being identical with the
Kappa distribution \citep{treu99a}.

\citet{Leub02} shows that Kappa-like distributions can result as a
consequence of the entropy generalization in nonextensive Tsallis
statistics \citep{Tsal95}, physically related to the long range
nature of the Coulomb potential, turbulence  and intermittency
\citep{Leub05, treu08}. The Kappa distribution is equivalent to the $q$ distribution function obtained from the maximisation of the Tsallis entropy. Systems subject to long-range interactions
and correlations are fundamentally related to non-Maxwellian
distributions \citep{Leub08}.   Core-halo distribution functions are
a natural equilibrium state in generalized thermostatistics
\citep{Leub04b}. Fundamental issues on Kappa distributions in space
plasmas and interplanetary proton distributions are emphasized in
\citet{Leub04}. \citet{Liva09} also examined how Kappa distributions
arise naturally from Tsallis statistical mechanics and provide a
solid theoretical basis for describing complex systems.

The generation of suprathermal electrons by resonant interaction
with whistler waves in the solar corona and wind was suggested by
\citet{Vock03, Vock08}. Introducing antisunward-propagating whistler waves into a kinetic model in
order to provide diffusion,  their results show that the whistler waves are capable of influencing the solar
wind electron VDFs significantly, leading to the formation of both the halo and strahl populations and a more isotropic
distribution at higher energies \citep{Vock05}.

 In an ambient quasi-static magnetic field,
plasma charges gain energy through the cyclotron resonance and the
transit time damping (magnetic Landau resonance) of the linear
waves. This is the case of high frequency whistler mode that enhance
the energy of electrons in Earth's foreshock \citep{ms98}, or that
of MHD waves which can accelerate both the electrons and the protons
in the solar flares \citep{mi91, mi97}, and in the inner
magnetosphere \citep{sm00}.

When large amplitude waves are present, the nonlinear Landau damping
can be responsible for the energization of plasma particles \citep{mi91, l00, sh07}.
Stochastic acceleration of plasma particles in compressional turbulence
seems to be consistent with the power law spectra occured throughout
the heliosheath downstream from the termination shock of the solar
wind \citep{fg06,fg07}. A mechanism for the
generation of  electron distribution function with suprathermal
tails in the upper regions of the solar atmosphere in the presence
of collisional damping was suggested by \citet{Vina00} as due to
finite-amplitude, low-frequency, obliquely propagating
electromagnetic waves. The nonthermal features of the VDFs can also
result from superdiffusion processes \citep{t97}, and due to heat
flows or the presence of the temperature anisotropies \citep{lv86}.

In the same spirit, \citet{Masu99} considered the steady state
solution of the Fokker-Planck (FP) equation and obtained a Kappa
distribution for a quasi-linear wave-particle diffusion coefficient
that varies inversely with the particle speed for velocities larger
than the thermal speed. \citet{sh07} used the same FP equation
to study the relative strengths of the wave-particle interactions
and Coulomb collisions. The formation of high-energy tails in the
electron VDF was also investigated with a FP model by
\citet{Lies97}.

Note that a one dimensional, electrostatic Vlasov model has
been proposed  for the generation of suprathermal electron tails in
solar wind conditions \citep{cali08}. The possible development of
Kappa velocity distribution was also illustrated in \citet{Hau07} by
the problem of low-frequency waves and instabilities in uniform
magnetized plasmas with bi-Maxwellian distribution.

Whatever the mechanisms of suprathermal tails formation, the kappa function is a useful mathematical tool to generalize the velocity distributions to the observed power law functions,  the particular Maxwellian VDF corresponding to the specific value of $\kappa \rightarrow \infty$.

\section{Theories based on the existence of Kappa distributions}
\subsection{Star's corona}
\citet{Scud92a, Scud92b} was pioneer to show the consequences of a
postulated nonthermal distribution in stellar atmospheres and especially
the effect  of the velocity filtration: the ratio of
suprathermal particles over thermal ones increases as a function of
altitude in an attraction field. The anticorrelation between the
temperature and the density of the plasma leads to this natural
explanation of velocity filtration for the heating of the corona,
without depositing wave or magnetic field energy. Scudder
\citep{Scud92b} determined  also the value of the kappa parameter
for different groups of stars. \citet{Scud94} showed that the excess
of Doppler line widths can also be a consequence of non thermal
distributions of absorbers and emitters. The excess brightness of
the hotter lines can satisfactorily be accounted for by a
two-Maxwellian electron distribution function \citep{Ralc07} and
should be also by a Kappa. Note that many solar observations
implicitely assume that the velocity distributions are Maxwellian in
their proper frame, so that the presence of suprathermal tails
should lead to reinterpretation of many observations.

 The mechanism of velocity filtration in solar corona
has been proposed  to explain the high
energy electrons at higher altitudes in the solar wind \citep{Scud92a, Scud92b}. Velocity
filtration was also applied  to heavy ions in the corona to explain
their temperatures more than proportional to their mass observed in the high
speed solar wind \citep{Pi03}.

Studying the heat flow carried by Kappa distributions in the solar
corona, \citet{Dore99} demonstrated that a weak power law tail in
the electron VDF can allow heat to flow up a radially directed
temperature gradient. This result was also confirmed by
\citet{Land01} who obtained the heat flux versus $\kappa$ in a slab of
the solar corona from a kinetic simulation  taking collisions into
account. For $\kappa >5$, the flux is close to the Spitzer-Harm
classical collisional values while for smaller values of $\kappa$,
the heat flux strongly increases and changes of sign! If $\kappa$ is
small enough, the fast wind can be suprathermally driven
\citep{zo05}.    This  shows the inadequacy of the classical heat
conduction law in space plasmas and the importance  to deal with
non-Maxwellian velocity distribution such as Kappa VDF
\citep{Meye99,Meye07}.

Note that the Kappa distribution is also consistent with mean
electron spectra producing hard X-ray emission in some coronal
sources \citep{kasp09}. Moreover, the low coronal electron temperatures and high ion charge states can be reconciled if the coronal electron distribution function starts to develop a significant suprathermal halo already below 3RS \citep{Esse00}. Effects of a Kappa distribution function of
electrons on incoherent scatter spectra were studied by
\citet{saito00}. The equilibrium ionization fractions of N, O, Ne,
Mg, S, Si, Ar, Ca, Fe and Ni were calculated for Maxwellian and
Kappa VDF based on a balance of ionization and recombination
processes \citep{wann03} for typical temperatures in astrophysical
plasmas. Low kappa values lead generally to a higher mean charge.

The Coulomb focusing effects on the bremsstrahlung spectrum are
investigated in aniso\-tropic bi-Lorentzian distribution plasmas in
\citet{Kim04}. Plasma screening effects on elastic electron--ion
collision processes in a Lorentzian (Kappa)-distribution plasma are
analyzed in \citet{jung00}.

\subsection{Solar wind}

\citet{PiLe96a} developed a kinetic model of
the ion-exospheres based on the Kappa VDF. The heat flux was
specified in \citet{PiLe98}. Their model has been applied to the
solar wind by \citet{Maks97b} and predicts the high speed solar wind
velocities with reasonable temperatures in the corona and without
additional acceleration mechanism.  The presence of suprathermal
electrons increases the electrostatic potential difference between
the solar corona and the interplanetary space and accelerates the
solar wind. The collisionless or weakly collisional models in corona
\citep{Scud92b, Maks97b, zo05}, all using VDFs with a suprathermal
tail, are able to reproduce the high speed streams of the fast solar
wind emitted out of coronal regions where the plasma temperature is
smaller, as well as the low speed solar wind originating in the
hotter equatorial regions of the solar corona.

\begin{figure}
\begin{center}
\includegraphics[width=11.5cm]{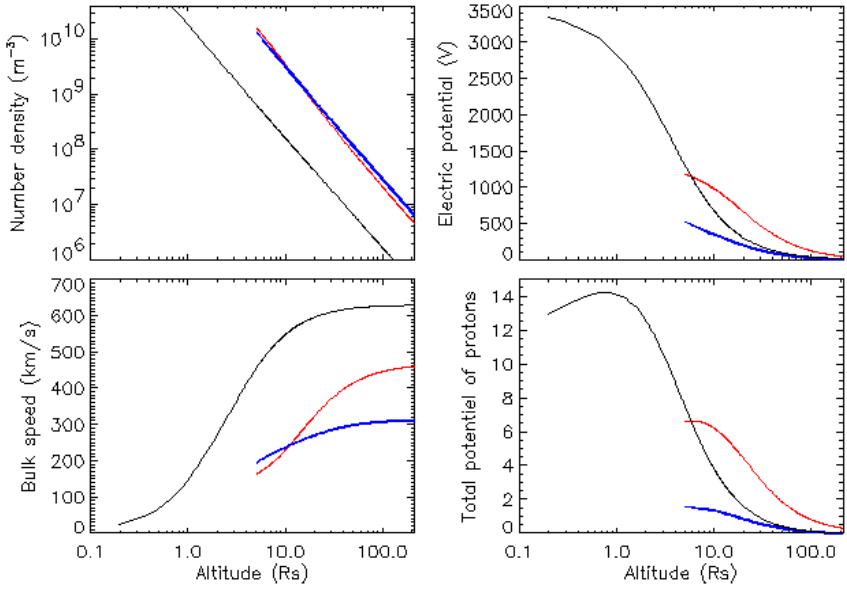}
\end{center}
\caption{Number density, electrostatic potential, bulk speed and
potential of protons for different solutions of the Kappa  kinetic
model of the solar wind. The lowest  blue velocity curve  (u=320
km/s at 215 Rs) in the bottom left panel corresponds to a Maxwellian
model with a starting radial distance (called exobase) at $r_0= 6$
Rs.  The middel red line (u=460 km/s at 215 Rs) corresponds to a
Kappa VDF with $\kappa=3$ and $r_0=6$ Rs. The upper black line
(u=640 km/s at 215 Rs) corresponds to a Kappa VDF for $\kappa=3$
with an exobase at $r_0=1.2$ Rs (Adapted from \citet{Lamy03b}).}
\label{kappavs}
\end{figure}

The exospheric
Lorentzian (or Kappa) model was extended to non monotonic
potential energy for the protons \citep{Lamy03} and shows that the
acceleration is especially large when it takes place at low radial
distances in the coronal holes where the number density is lower
than in other regions of the corona, as illustrated on Fig.
\ref{kappavs}. \citet{zoug04} demonstrated with a parametric study
that this acceleration is a robust result produced by the presence
of a sufficient number of suprathermal electrons and is valid also for
other VDF with suprathermal tails than Kappa.

The acceleration of the solar wind heavy ions is investigated in \citet{PiLa04}: due to their different
masses and charges, the minor ions reach different velocities. Even if their mass on charge ratio is always larger than that of the protons, they can be accelerated to velocities larger
than that of the protons if their temperatures are sufficiently high
in the corona.

Adding the effects of the Coulomb collisions, a kinetic solar wind
model based on the solution of the Fokker-Planck equation was
developed \citep{pml99, pml01}. Typical electron VDFs measured at 1 AU
by WIND have been used as boundary condition to determine the VDFs
at lower altitudes and it was proved that, for several solar
radius, the suprathermal populations must be present in the corona
as well \citep{pml99}. Indeed, since the particle free path
increases as $v^{4}$ in a plasma due to the properties of Coulomb
collisions, the suprathermal particles are non collisional even when
thermal particles are submitted to collisions. High energy tails can
develop for Knudsen numbers (i.e. ratio of mean free path over scale
height) as low as $10^{-3}$ \citep{Shou83}. \citet{ml85} have
studied the collisional relaxation process and the associated rates
(diffusion and friction) for a nonthermal solar wind with Kappa
VDFs. The Fokker-Planck model \citep{pml01} illustrated also the
transformation of the VDF of the electrons in the transition region
between the collision-dominated region in the corona and the
collisionless region at larger radial distances. The VDF became more
and more anisotropic in the transition region. Contrary to
exospheric models that are analytic, these collisional kinetic
models have to be solved numerically. The spectral method of
expansion of the solution in orthogonal polynomials converges faster
for suprathermal plasmas by using polynomials based on the Kappa
function weight developed by \citet{MaPi08}.

As a result of low collision rates in the interplanetary plasma, the
electrons and the ions develop temperature anisotropies and
their VDFs become skewed and develop tails and heat fluxes along the
ambient magnetic field \citep{ma82, pi87, sa03, st08}. Moreover,
field-aligned fluxes of (suprathermal) particles can be encountered
at any altitude in the solar wind \citep{pi90}, but they become
prominent in energetic shocks, like the coronal mass ejections or
the fast solar wind at the planetary bow shock, giving rise to
counterstreaming plasma events \citep{fe74,go93,Stei05,gf06}. The same
mechanisms mentioned above can induce anisotropic nonthermal VDFs in
counterstreaming plasmas, and such complex plasmas can hold an
important amount of free energy that makes them unstable to the
excitation of waves and instabilities.

\citet{fain96} suggested that the non-thermal electrons can
contribute as much as 50 \% of the total electron pressure within
magnetic clouds. Using this hypothesis, \citet{Niev08} have assumed
two populations of electrons in magnetic clouds: a core Maxwellian
and a halo Kappa-like distribution. They found that kappa values
exhibit either minor differences or, for some events, can be greater
inside than outside the magnetic clouds.

Observations from VOYAGER indicate that ions in the outer heliosphere are well described by Kappa functions \citep{Deck05}.
The effects of a Kappa distribution in the heliosheath on the global
heliosphere and energetic neutral atoms (ENA) flux have been
studied in \citet{heer09}. The use of Kappa, as opposed to a
Maxwellian, gives rise to greatly increased ENA fluxes above 1 keV,
while medium energy fluxes are somewhat reduced. The effect of a
Kappa distribution on the global interaction between the solar wind
and the local interstellar medium (LISM) is generally an increase in
energy transport from the heliosphere into the LISM, due to the
modified profile of ENA's energies. This results in a motion of the
termination shock (by 4 AU), of the heliopause (by 9 AU) and of the
bow shock (25 AU) farther out, in the nose direction.

\subsection{Earth's exosphere}

The Kappa model of ion-exosphere \citep{PiLe96b} has been used to
study different plasma regions in the magnetosphere of the Earth.
Pierrard \citep{Pier96, PiKa07} obtained new current-voltage
relationships in auroral regions when suprathermal particles are
assumed to be present with Lorentzian and Bi-Lorentzian
distributions. Field-aligned conductance values were also estimated
from Maxwellian and Kappa distributions in quiet and disturbed
events using Freja electron data \citep{Olss98}.

Introducing a Kappa model appears to resolve discrepancy between
calculations and observations of resonant plasma echoes and emissions used for
{\it in-situ} measuring the local electron density and the magnetic field strength
in the magnetospheric environments \citep{vm05}.

The three dimensional plasmasphere has been modeled  using
Kappa velocity distribution functions for the particles \citep{PiSt08}: this  physical dynamic
model of the plasmasphere gives the position of the plasmapause and the number density of the particles inside and outside the plasmasphere. The effects of suprathermal particles on the temperature in the
terrestrial plasmasphere were illustrated using Kappa functions in
\citet{PiLe01}.

The terrestrial polar wind is in some way similar to the
escape of the solar wind: similar effects of suprathermal particles appear and
lead to an increase of the escaping flux \citep{LePi01, Tam07}.
Along open magnetic field lines, the wind speed is increased by the
presence of suprathermal particles.  A Monte Carlo simulation developed by \citet{barg01} shows the
transformation of $\rm H^+$ polar wind velocity distributions  with Kappa suprathermal
tails in the collisional transition region.

\subsection{Planetary exospheres}

\citet{Meye01} emphasized the importance of not being Maxwellian for
the large structure of planetary environments. For bound structures
shaped along magnetic field lines, the temperature increases with
the distance, in contrast to classical isothermal equilibrium (see
Table 1). The rise in temperature as the density falls is a generic
property of distributions with suprathermal tails,  as shown in
\citet{Meye95} and \citet{Monc02} to explain the temperature inversion in the
Io torus.

The ion-exosphere Kappa model \citep{PiLe96a} has been adapted to the Saturnian
plasmasphere \citep{Moor05}. The kappa index gives an additional
parameter to fit observations of Cassini. The polar wind and
plasmasphere of Jupiter and Saturn were also recently modeled with
Kappa functions \citep{Pier09}: the suprathermal particles increase
significantly the escape flux from these giant planets, so that the
ionosphere become an important source for their inner magnetosphere.

Spacecraft charging environments at the Earth, Jupiter and Saturn
were also obtained by \citet{Garr00} using Kappa distributions for
the warm electrons and protons.

\section{Dispersion properties and stability of Kappa distributions}

In many circumstances, the wave-particle interactions can be made
responsible for establishing non-Maxwellian particle distribution
functions with an enhanced high energy tail and shoulder in the
profile of the distribution function. In turn, the general plasma
dynamics and dispersion properties are also altered by the presence
of nonthermal populations. Thus, the waves and instabilities in
Kappa distributed plasmas, where collisions are sufficiently rare,
are investigated using kinetic approaches based on Vlasov-Maxwell
equations.

\subsection{Vlasov-Maxwell kinetics. Dielectric tensor}

Using a kinetic approach, \citet{st94} have calculated the
dielectric tensor for the linear waves propagating at an arbitrary
angle to a uniform magnetic field in a hot plasma with particles
modeled by a Kappa distribution function. Despite the fact that the
elements of this tensor take complicate integral forms, this paper
is of reference for the theory of waves in Kappa-distributed
plasmas. This dielectric tensor can be applied for analyzing the
plasma modes as well as the kinetic instabilities in a very general
context, limited only by the assumptions of linear plasma theory.
The analytical dispersion relations derived previously \citep{ts86}
in terms of the modified Bessel functions of the lowest order $I_0$,
$I_1$, $K_0$, and $K_1$ (which are tabulated) were restricted to a
weak damping or growth of plasma waves by resonant interactions with
plasma particles.

The approach developed in \citet{st94} also restricts to the
distributions functions which are even functions of parallel
velocity of particles, $v_{\parallel}$ (where parallel or
perpendicular directions are taken with respect to the stationary
magnetic field), and this is what the authors called an usual
condition in practice. This condition fails only in some extreme
situations \citep{st94} as are, for instance, the asymmetric
beam-plasma structures developing in astrophysical jets or more
violent shocks. These cases can however be approached distinctively
in the limits of waves propagating along or perpendicular to the
ambient magnetic field \citep{lspt08, ltsp09, laza09}.


\subsection{The modified plasma dispersion function}

Early dispersion studies have indeed described the simple
unmagnetized plasma modes and the field aligned waves in the
presence of an ambient magnetic field \citep{l83,st91}. Let us first
remember the most important analytical changes introduced by the
power law distributions of plasma particles.

Kinetic theory of an equilibrium Maxwellian plasma naturally produce
a dispersion approach based upon the well-know Fried and Conte
plasma dispersion function \citep{fc61}
\be Z (f) = {1 \over \pi^{1/2}} \int_{-\infty}^\infty dx \,
{\exp(-x^2) \over x-f}, \; \;\; f = {\omega \over k} \sqrt{m \over
2k_BT}, \; \;\; {\rm Im}(f) > 0. \label{e2} \ee
For a non-Maxwellian plasma characterized by the Kappa distribution
function (\ref{e1}), \citet{st91} have derived a modified plasma
dispersion function
\be Z_{\kappa}(f) = {1 \over \pi^{1/2} \kappa^{1/2}} \, {\Gamma
(\kappa) \over \Gamma \left( \kappa -{1 \over 2}\right)} \,
\int_{-\infty}^{+\infty} dx \, {(1+x^2/\kappa)^{-(\kappa +1)} \over
x - f}, \; \;\; {\rm Im}(f) > 0 , \label{e3} \ee
where the spectral index $\kappa$ was first restricted to positive
integers $\kappa > 1.5$. This new dispersion function has been
ingenuously generalized to arbitrary real $\kappa > 1.5$ by
\citet{mh95} also showing that it is proportional to the Gauss
hypergeometric function, $_2F_1$. Extensive characterizations for
this Kappa dispersion function have been made by \citet{st91, mh95,
mace03, vd07, mh09}. Furthermore, for a (space) plasma immersed in a
stationary magnetic field, which determines a preferred direction
for the acceleration and motion of electrons and ions, \citet{hm02}
have introduced an hybrid Kappa-Maxwellian distribution function,
one-dimensional Kappa along the magnetic field lines and Maxwellian
perpendicular to this direction. The new dispersion function
$Z_{\kappa M}$ obtained for this anisotropic plasma differs from
$Z_{\kappa}$ in (\ref{e3}) in the power to which the term
$(1+x^2/k)^{-1}$ is raised: in the isotropic case it is $\kappa +
1$, while in the one-dimensional case it is $\kappa$. Relation
between these two integral functions takes two forms \citep{hm02,
lss08}
\begin{align}  Z_{\kappa M}(f) & = {(\kappa -1)^{3/2} \over (\kappa
-3/2)\kappa^{1/2}} Z_{\kappa-1}\left[\left({\kappa -1 \over
\kappa}\right)^{1/2}f \right] \notag \\ & = \left(1+{f^2 \over
\kappa}\right) \, Z_{\kappa}(f) + {f \over \kappa} \, \left(1-{1
\over 2\kappa} \right), \label{e4} \end{align}
and both $Z_{\kappa M}(f)$ and $Z_{\kappa}(f)$ functions approach
the Maxwellian dispersion function $Z(f)$ from (\ref{e2}) in the
limit of $\kappa \to \infty$.

\subsection{Isotropic Kappa distributions}

The effect of an isotropic Kappa population on plasma modes has been
described by the first dispersion studies already reviewed by
\citet{hmv00} and \citet{hell06}.

\subsubsection{Unmagnetized plasma}

The fluctuations are, in general, enhanced in low-$\kappa$ plasmas,
and the electromagnetic and electrostatic dispersion relations show
significant dependence on the spectral index $\kappa$ \citep{ts91,
st92, mh95, mht98}.

For  Langmuir oscillations, Landau damping in a hot, isotropic,
unmagnetized plasma is controlled by the spectral index while for
ion-acoustic waves Landau damping is more sensitive to the ion
temperature than the spectral index \citep{Qure06}. Thus, Landau
damping growth rates of long wavelengths Langmuir modes become much
larger for a Lorentzian (Kappa) plasma \citep{ts91} limiting the
existence of the (weakly damped) Langmuir waves in space plasmas
with Kappa distributions, to a narrow-band just above the electron
plasma frequency. The significant increase of spatial Landau damping
by small $\kappa$ electrons is also demonstrated for spatially
damped plasma waves generated by a planar grid electrode with an
applied time harmonic potential \citep{pod05}. Hybrid models of
Maxwellian plasmas partially populated by hot $\kappa$-components
can provide better fits to the observations and experiments than the
simple uniform Maxwellian \citep{hel00,mace99}.

The relativistic effects on dispersion and Landau damping of
Langmuir waves in a relativistic Kappa-distributed plasma have been
studied by \citet{pod08}. The relativistic dispersion relations
derived have been used to compute the damping rates and phase speeds
for plasma waves in the solar wind near the Earth orbit. It was
found a good match for the electron velocities in the superhalo with
the phase speed of weakly damped plasma waves, and thus providing a
plausible mechanism for their acceleration.

The existence conditions and characteristics of ion-acoustic
solitary waves have been studied by \citet{Abba08,saini09} and
\citet{chu09} showing that Kappa-distributed electrons are not
favorable to the existence of these solitons. A comparative study of
Langmuir waves, dust ion acoustic waves, and dust-acoustic waves in
Maxwellian and Kappa distributed plasmas is presented in
\citet{zah04}. Landau damping rate of dust acoustic waves in a dusty
plasma modeled by a Kappa distribution for electrons and ions and by
a Maxwellian for the dust grains has been found to be dependent on
the spectral index $\kappa$ as well as the ratio of ion density to
electron \citep{Lee07}. Dust acoustic solitons have also been
studied in plasmas with Kappa-distributed electrons and/or ions and
cold negative or positive dust grains \citep{Balu08, you08} or with
nonthermal ions having Kappa-vortex-like velocity distributions
functions \citep{kam09}.

\subsubsection{Magnetized plasmas}

The general dielectric tensor for magnetoplasmas comprising
components with generalized Lorentzian distributions has been
calculated by \citet{st94} for arbitrary oriented wave-vectors.
Applying to the electrostatic or electromagnetic waves propagating
parallel to the ambient magnetic field, simple dispersion relations
can be derived \citep{st91,xue93, st94,mace96, m98} in terms of the
modified plasma dispersion function (\ref{e3}).

This dielectric tensor has also been simplified for a 3-dimensional
isotropic Kappa distribution in a form similar to that obtained by
Trubnikov \citep{mace96}, and for a Kappa loss-cone distribution
with applications to a large variety of space plasmas like the solar
wind, magnetosheath, ring current plasma, and the magnetospheres of
other planets \citep{mace96,xts98,Pokh02,xiao06,xial06,sing07}.

For wave-vectors oblique to the magnetic field,
\citet{mace03,mace04} has described the generalized electron
Bernstein modes in a plasma with an isotropic Kappa velocity
distribution. In a hybrid Kappa-Maxwellian plasma, unlike the
uniform Maxwellian plasma, the dispersion properties of the oblique
electromagnetic waves were found to be markedly changed from an
elaborate study including effects of the Kappa value, the
propagation angle, and the temperature anisotropy on dispersion and
damping \citep{cat07}. Notice that, because of the anisotropy of the
countours in the velocity space, such a Kappa-Maxwellian
distribution is unstable in an overdense plasma near the
electron-cyclotron frequency even when the parallel and
perpendicular temperatures are equal.

The dispersion relations for low-frequency hydromagnetic waves in a
Kappa distributed plasma has recently been derived by \citet{bas09}
showing that both the Landau damping and the transit-time damping
(magnetic analogue of Landau damping) of the waves are enhanced in
the suprathermal region of the velocity space.

\subsection{Anisotropic Kappa distributions}

The effects of anisotropic Kappa distributions have also been
investigated since the anisotropic velocity distributions are most
probably at the origin of nonthermal emissions in astrophysical
sources, and the magnetic field fluctuations in space plasma.

\begin{figure}
\begin{center}
\includegraphics[width=5.5cm]{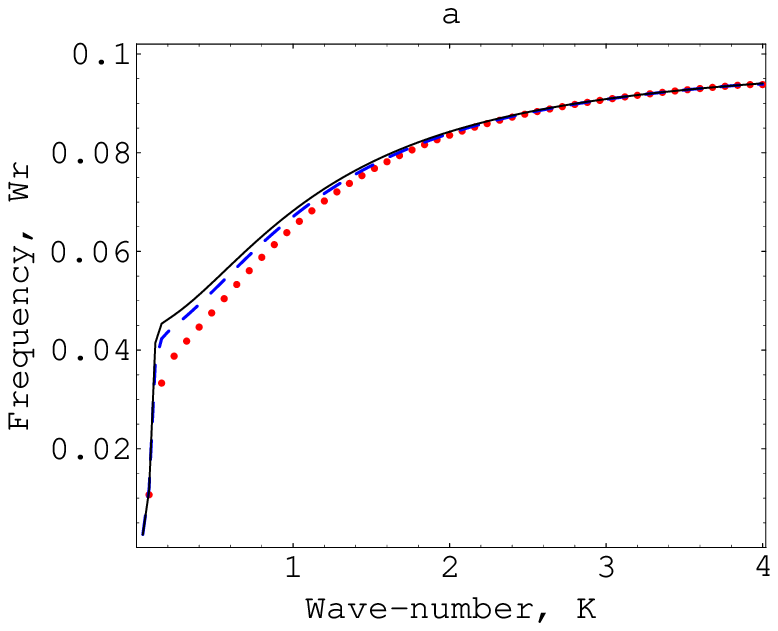} $\;\;\;$
\includegraphics[width=5.5cm]{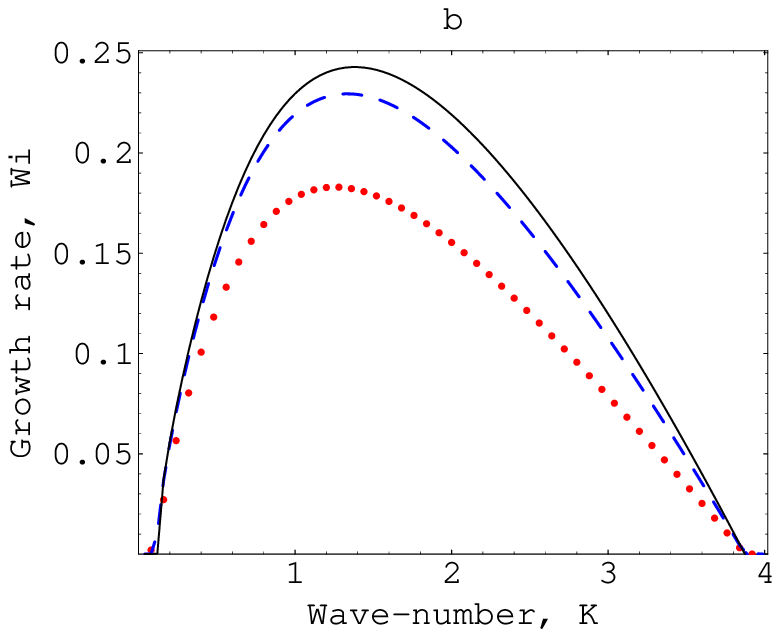}
\end{center}
\caption{Dependence of dispersion curves (panel a) and growth rates
(b) for the electron cyclotron instability (R mode) on the spectral
index $\kappa =$ 2 (red dotted lines), 4 (blue dashed lines), and
$\kappa \to \infty$ for Maxwellian plasmas (black solid lines).
These are exact numerical solutions for parameters intended to
simulate (low altitude) solar wind conditions: plasma temperature
$T_e =2.5 \times 10^7$ K, a temperature anisotropy $T_{\perp}/T_e
\simeq 4 $, and gyrofrequency $|\Omega_e| = 0.1 \omega_{pe}$. The
coordinates are scaled as Wr$=\omega_r /\omega_{pe}$ Wi$=\omega_i
/\omega_{pe}$ and K = $kc /\omega_{pe}$. (Adapted from
\citet{lspt08}.)} \label{fig4}
\end{figure}

When both electrons and ions are modeled by an anisotropic
distribution of bi-Kappa type, \citet{st91} have derived the general
dispersion relation for the parallel electromagnetic  modes,
right-handed (R mode) and left-handed (L mode) circularly polarized,
in terms of the modified plasma dispersion function (\ref{e3}). The
effect of suprathermal particles on the stability of these modes
largely varies depending on the shape of the distribution function,
and the mode frequency, whether it fits to thermal Doppler shift of
the electron gyrofrequency (the whistler instability driven by the
cyclotron resonance with electrons) or the ion gyrofrequency (the
electromagnetic ion cyclotron instability). Thus, while the growth
rates of the electron cyclotron instability (R-mode lower branch,
$\omega_r \leq | \Omega_e |$) become lower in a bi-Kappa plasma than
for a bi-Maxwellian with the same temperature anisotropy
\citep{lspt08}, also see Fig. \ref{fig4}, b, at smaller frequencies
($\omega_r \ll |\Omega_e|$) , the whistler growth rates become
higher \citep{m98, trip08}. \citet{cat07} have also shown that,
unlike a bi-Maxwellian plasma, the low-frequency whistler modes in a
Maxwellian-Kappa plasma (described above), can be stable to the
temperature anisotropy. Their study includes effects of varying
$\kappa$ for both underdense and overdense plasmas, and for both
parallel and oblique propagation.

The loss cone bi-Lorentzian distribution, which allows to plasma
populations to have anisotropic temperatures and a loss cone, has
been used extensively (although the latter is largely
inconsequential in models for wave propagation parallel to the
magnetic field) in space and laboratory applications \citep{m98,
trip08, trip00}. \citet{sing07} obtained the components of the
dielectric tensor for such a distribution function and made
parametric studies of the effect of $\kappa$-index, the loss-cone
index (which is, in general different from $\kappa$), and different
temperature anisotropies \citep{trip08}. Whistler mode instability
in a Lorentzian (Kappa) magnetoplasma in the presence of
perpendicular AC electric field and cold plasma injection was
studied by \citet{trip00}. An unperturbed Lorentzian distribution
has also been used for studying the effect of a cold plasma beam on
the electromagnetic whistler wave in the presence of a perpendicular
AC electric field in the Earth's atmosphere \citep{Pand08}.

The Kappa-loss-cone (KLC) distribution function obeys a power-law
not only at the lower energies but also at the relativistic
energies. A relativistic KLC distribution has been introduced by
\citet{xiao06} for an appropriate characterization of the energetic
particles found in planetary magnetospheres and other plasmas, where
mirror geometries occur, i.e., a pronounced high energy tail and an
anisotropy. The field-aligned whistler growing modes in space
plasmas have been investigated by \citet{xial06} and \citet{zhou09}
applying relativistic treatments for relativistic Kappa or KLC
distributions. The threshold conditions for the whistler instability
in a Kappa distributed plasma have been derived by \citet{xiaojgr}.
Numerical calculations were carried out for a direct comparison
between a KLC distribution and the current Kappa distribution. The
KLC was also adopted to model the observed spectra of solar
energetic protons \citep{xiaoal08b}. Recent studies \citep{xiaoal08,
zhou09} have introduced a generalized relativistic Kappa
distribution which incorporates either temperature anisotropy or
both loss cone and temperature anisotropy.

Purely growing ($\omega \simeq 0$) mirror modes and low frequency
electromagnetic ion cyclotron waves are widely detected in the solar
wind and magnetosheath plasmas being driven by an ion temperature
anisotropy, $T_{\perp} > T_{\parallel}$ (or pressure, $p_{\perp} >
p_{\parallel}$). The effects of the suprathermal tails on the
threshold conditions and the linear growth rates of these
instabilities have widely been investigated in the last decade
\citep{Leubsc00, Leubsc01, Leubsc02, Ged01, Pokh02}. An universal
mirror wave-mode threshold condition for non-thermal space plasma
environments was obtained by \citet{Leubsc00, Leubsc01, Leubsc02}.
The linear theory of the mirror instability in non-Maxwellian space
plasmas was developed by \citet{Pokh02} for a large class of Kappa
to suprathermal loss cone distributions in view of application to a
variety of space plasmas like the solar wind, magnetosheath, ring
current plasma, and the magnetospheres of other planets. Thus, while
transition to nonthermal features provides a strong source for the
generation of mirror wave mode activity, reducing drastically the
instability threshold \citep{Leubsc02}, a more realistic presence of
suprathermal tails exclusively along the magnetic field (parallel
$\kappa$-distribution) counteracts the growth of the mirror
instability and contributes to stabilization \citep{Pokh02}. In the
nonlinear regime, solitary structures of the mirror waves occur with
the shape of magnetic holes \citep{Pokh08} suggesting that the main
nonlinear mechanism responsible for mirror instability saturation
might be the magnetic trapping of plasma particles.

\begin{figure}
\begin{center}
\includegraphics[width=5.cm]{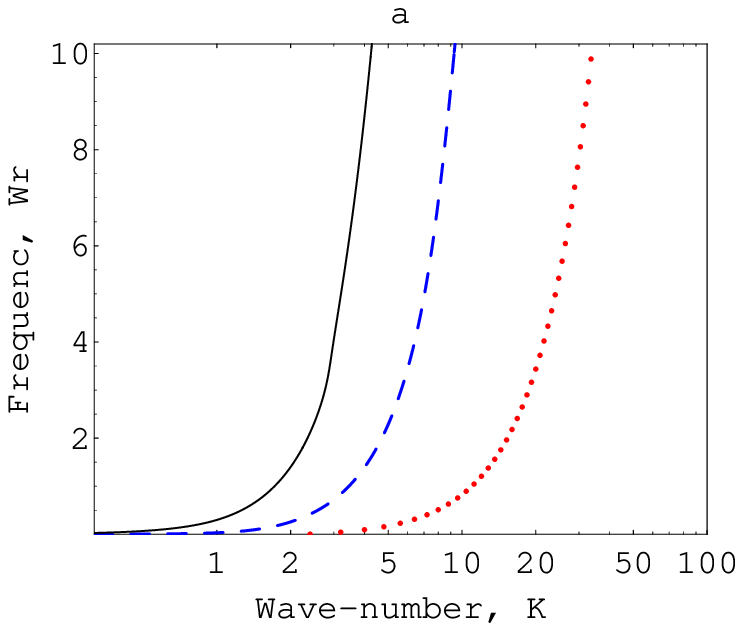} $\;\;\;$
\includegraphics[width=5.cm]{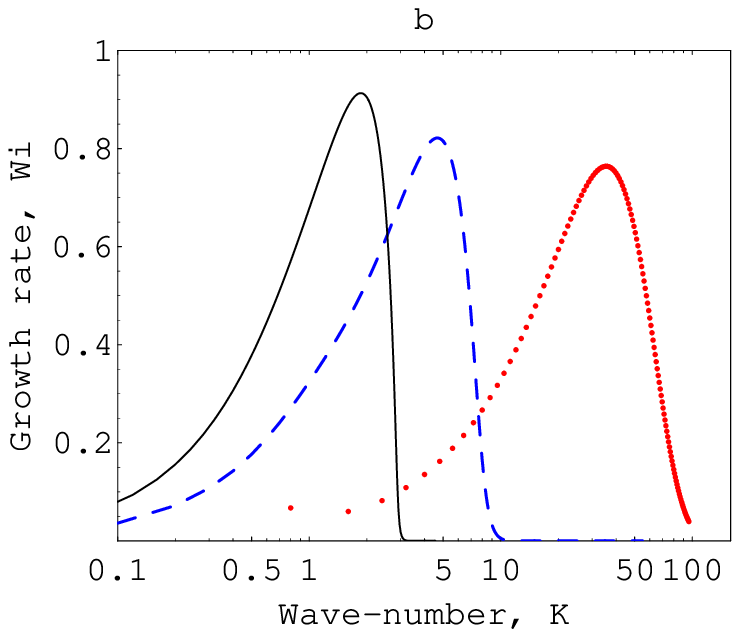}
\end{center}
\caption{Dependence of dispersion curves (a) and growth rates (b)
for firehose instability on the spectral index $\kappa =$ 4 (red
dotted lines), 5 (blue dashed lines), and $\kappa \to \infty$ for
Maxwellian plasmas (black solid lines). These are numerical exact
solutions obtained with parameters estimated for solar flares:
plasma density $n= 5 \times 10^{10}$ cm$^{-3}$, electron and proton
temperature $T_{e,\perp} = T_{p,\perp} = T_{p, \parallel} \simeq
10^7$ K, the electron temperature anisotropy
$T_{e,\perp}/T_{e,\parallel} \simeq 20 $ (in a), and $\perp$ and
$\parallel$ denote directions with respect to the ambient magnetic
field $B_0 = 100$ G. The coordinates are scaled as Wr$=\omega_r
/\Omega_{p}$, Wi$=\omega_i /\Omega_{p}$ and K = $kc /\omega_p$.
(Adapted from \citet{laza09})} \label{fig5}
\end{figure}

The ion cyclotron wave instability driven by a temperature
anisotropy ($T_{\perp}/T_{\parallel} > 1$) of suprathermal ions
(protons) modeled with a typical Kappa distribution is investigated
by \citet{xue93, xiao07,xue96a, xue96b} for solar wind conditions
and for magnetosphere. The threshold condition for this instability
is determined by \citet{xiao07b}. As the spectral index $\kappa$ for
protons increases, the maximum growth rates of R, L modes decrease
\citep{st92,xue93,das03,xue96b} and the instability threshold
generally decreases and tends to the lowest limiting values of the
bi-Maxwellian ($\kappa \to \infty$) \citep{xiao07b}. The
corresponding enhancement in the growth rate of L mode waves in
planetary magnetospheres is less dramatic, but the Kappa
distribution tends to produce a significant wave amplification over
a broader range of frequency than a Maxwellian distribution with
comparable bulk properties \citep{xue93}. The damping and growth
rates of oblique waves are also lower for Kappa distributions, but
differences become less important for nearly perpendicular waves
\citep{xue96a}.

\begin{figure}
\begin{center}
\includegraphics[width=5.5cm]{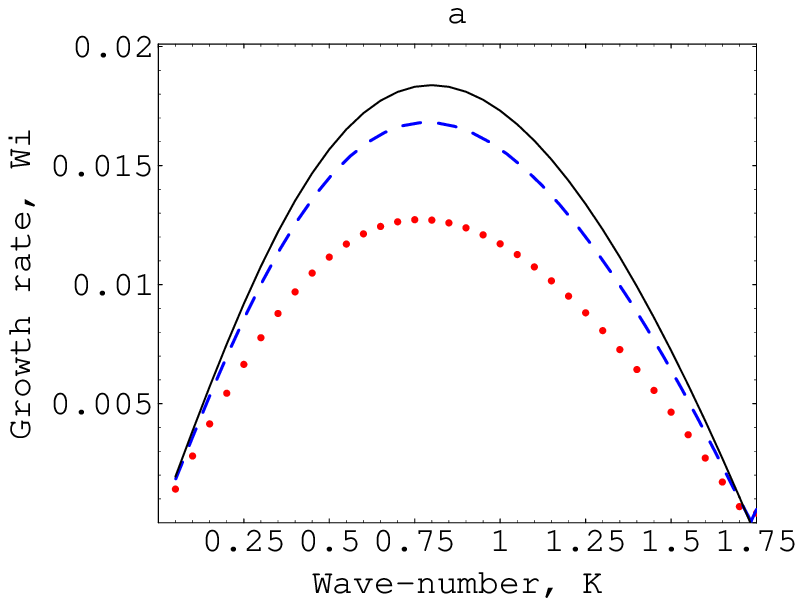} $\;\;\;$
\includegraphics[width=5.5cm]{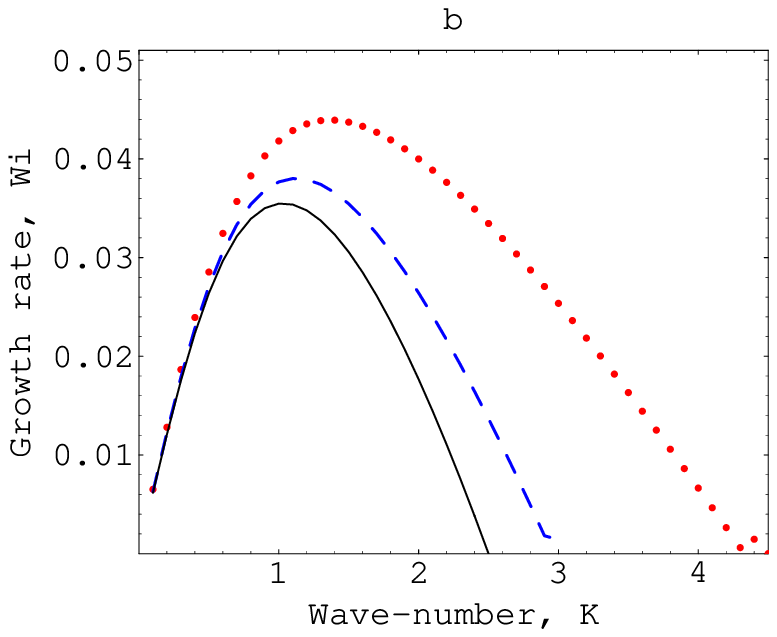}\\
\includegraphics[width=5.5cm]{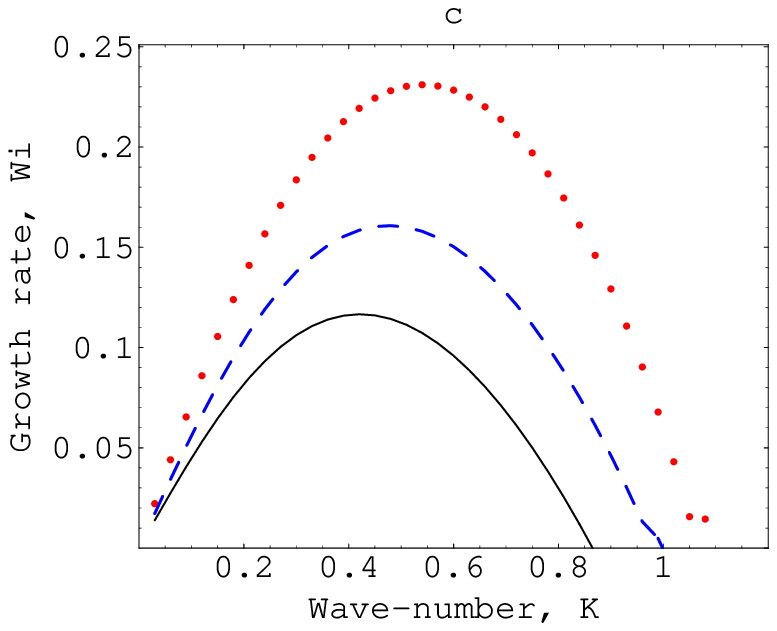}
\end{center}
\caption{Dependence of growth rates for Weibel instability (panel
a), filamentation (b) and two stream instability (c), on the
spectral index $\kappa =$ 2 (red dotted lines), 4 (blue dashed
lines), and $\kappa \to \infty$ for Maxwellian plasmas (black solid
lines). Here the parameters are typical for solar wind conditions:
plasma temperature $T_e \simeq 2 \times 10^6$ K, a temperature
anisotropy $T_{\perp}/T_e \simeq 4 $ (in a), and symmetric
counterstreams in b and c, with the same temperature and the bulk
velocity $v_0 = 0.1 c$ (where $c = 3 \times 10^8$ m$/$s is the speed
of light in vacuum). The coordinates are scaled as Wi=$\omega_i
/\omega_{pe}$ and K = $kc /\omega_{pe}$ in panels a and b, and K =
$kv_0 /\omega_{pe}$ in c. (Adapted from \cite{lss08, lspt08}.)}
\label{fig6}
\end{figure}

In the opposite case, a surplus of parallel kinetic energy,
$T_{\perp}< T_{\parallel}$ (or pressure $p_{\perp} < p_{\parallel}$)
will excite other two kinetic instabilities: the firehose
instability propagating parallel to the magnetic field lines and
with maximum growth rates of the order of ion gyrofrequency
\citep{laza09}, and the Weibel like instability propagating
perpendicular to the hotter direction (in this case also
perpendicular to the magnetic field lines), and which, in general,
is much faster reaching maximum growth rates of the order of
electron (or ion) plasma frequency \citep{lss08, ltsp09}. Recent
investigations \citep{laza09} of the electron firehose instability
driven by an anisotropic electron distribution of bi-Kappa type,
have proven that, by comparison to a bi-Mawellian, the threshold
increases, the maximum growth rates are slightly diminished and the
instability extends to large wave-numbers (see Fig. \ref{fig5}).
Instead, a more important reduction has been found for the growth
rates of the electron Weibel instability (Fig. \ref{fig6}, a) in a
non-magnetized or weakly magnetized plasma with bi-Kappa
distributions \citep{zm07,lss08,lspt08,ltsp09}.

Despite the similar features of the electromagnetic filamentation
instability, which is a Weibel-like instability that grows in a
counterstreaming plasma or a beam-plasma system perpendicular to the
streaming direction, the effect of suprathermal populations is
opposite enhancing the filamentation growth rates (Fig. \ref{fig6},
b) \citep{lss08, ltsp09}. Extended investigations have included
counterstreaming plasmas with an internal bi-Kappa distribution,
when the filamentation and Weibel effects cumulate leading again to
increased growth rates but only for plasmas hotter in the streaming
direction. Otherwise, if counterstreaming plasmas are hotter in
perpendicular direction, the effective anisotropy decreases,
diminishing the growth rates of filamentation instability
\citep{lss08,ltsp09}. However, in this case, two other unstable
modes are expected to arise, both along the streams: a Weibel-like
electromagnetic instability and a two-stream electrostatic
instability, which is, in general, faster than Weibel. Furthermore,
suprathermal populations enhance the electrostatic instability
leading to larger growth rates for lower $\kappa$ (Fig. \ref{fig6},
c). The same behavior has been observed for the modified two-stream
instability driven by the relative motion of ions, assumed
Kappa-distributed, with respect to the electrons (assumed
Maxwellian): maximum growth rates decrease with $\kappa$ and with
plasma $\beta$ \citep{lang05}.

Recently, \citet{bas08, bas09} provided a systematic study for the
stability of a magnetized plasma at low frequencies and in various
limits of low or high plasma beta (plasma pressure/magnetic
pressure), showing that the threshold values for the excitation of
the unstable hydromagnetic waves in Kappa distribution plasma are
increased as a consequence of enhancing the resonant wave-particle
damping.

\section{Outlines and perspectives of Kappa distributions}

The Kappa function has been proved to be a convenient tool to
describe plasma systems out of thermodynamic equilibrium. Since
the fast particles are nearly collisionless in space plasmas,
they are easily accelerated and tend to produce nonequilibrium
velocity distributions functions with suprathermal tails decreasing
as a power law of the velocity. Thus, models based on the kappa
distribution allow to analyze the effects of the suprathermal
particles and to fit distributions for ions and electrons measured
in situ, in the solar wind and in the magnetosphere of the
planets.

Major consequences follow the presence of these suprathermal
particles, and especially the velocity filtration which makes the
kinetic temperature increase upwards. This mechanism has been
proposed to explain the heating of the corona. The suprathermal
particles increase also the escape flux in planetary and stellar
wind and can explain the acceleration of the fast solar wind.  For
low values of $\kappa$, the heat flux changes of sign compared to Spitzer-Harm so that heat
can flow from cold to hot.  The future solar missions (Solar Orbiter
and Solar Probe) should improve the observations concerning the presence of suprathermal particles in the corona.

Valuable theories have been proposed concerning the origin and the
fundamental physical arguments for a kappa family distribution
function. The universal character of these distribution functions
suggests that they can be attributed to a particular thermodynamic
equilibrium state related to the long range properties of Coulomb
collisions. Thus, Kappa distributions generalizes the notion of
equilibrium for collisionless plasmas far from thermal
(Boltzmann-Maxwell) equilibrium, but containing fully developed
quasistationary turbulent fields. A new statistical mechanical
theory has been proposed extending Gibbsian theory and relaxing the
independence of subsystems (no binary collisions) through
introducing a generalized Kappa function dependence on entropy. In
the absence of binary collisions, the Kappa parameter has been found
to be a measure for the strength of the subsystem correlations
introduced by the turbulent field fluctuations. Particular forms for
a nonextensive (superadditive) entropy and for temperature have been
derived to satisfy the fundamental thermodynamic relations and
recover exactly the family of Kappa distributions from (\ref{e1}).

An isotropic Kappa distribution will therefore be stable against the
excitation of plasma instabilities. This function replaces the
Boltzmann-Maxwell distribution in correlated collisionless
equilibria of plasma particles and turbulent fields.
But such dilute plasmas easily develop flows and temperature
anisotropies making Kappa distributions to deviate from isotropy and
drive kinetic instabilities. In space plasmas embeded by the
interplanetary magnetic field, heating and instability are, in
general, resonant, and both will be enhanced in Kappa distributed
plasmas because there are more particles available at high energies
to resonate with waves. Revisiting transport theories and
calculation of the transport coefficients in solar environments is
therefore an important task, and a significant progress is expected
to be done involving Kappa distributions.

Many solar models have been developed on a questionable existence of
a Maxwellian equilibrium, so that, now, the presence of Kappa
distributions asks for more realistic reinterpretations that are
expected to provide better fits to the observations. The new
techniques developed for measuring the electron density,
temperature, and the suprathermal parameters will also offer
important clues to understanding transport properties in space plasmas.

\begin{acknowledgements}
The authors acknowledge support from the Belgian Institute for Space 
Aeronomy (VP), and from the Katholieke Universiteit Leuven and
the Ruhr-Universit\" at Bochum - Research Department Plasmas with Complex 
Interactions (ML).
\end{acknowledgements}



\begin{thebibliography}{}
\bibitem[Abbasi and Pajouh (2008)]{Abba08}Abbasi, H., and
Pajouh, H.\,H.: 2008, \emph{Phys. Plasmas Control. Fus.} {\bf 50}, 095007.
\bibitem[Baluku and Hellberg (2008)]{Balu08}Baluku, T.\,K., and
Hellberg, M.\,A.: 2008, \emph{Phys Plasmas} {\bf 15}, 123705.
\bibitem[Bame et al. (1967)]{Bame67}Bame, S.\,J., Asbridge, J.\,R.,
Felthauser, H.\,E., Hones, E.\,W., Strong, I.\,B.: 1967, \emph{J.
Geophys. Res.} {\bf 72}, 113-129
\bibitem[Barghouthi et al. (2001)]{barg01}Barghouthi, I., Pierrard,
V., Barakat A.\,R., and Lemaire, J.: 2001, \emph{Astrophys. Space
Sci.} {\bf 277}, 427.
\bibitem[Basu (2008)]{bas08}Basu, B.: 2008, \emph{Phys. Plasmas} {\bf 15}, 042108.
\bibitem[Basu (2009)]{bas09}Basu, B.: 2009, \emph{Phys. Plasmas} {\bf 16}, 052106.

\bibitem[Califano and Mangeney (2008)]{cali08}Califano, F., and
Mangeney, A.: 2008, \emph{J. Geophys. Res.} {\bf 113}, A06103.
\bibitem[Cary et al. (1981)]{ca81}Cary, J.\,R., Thode, L.\,E., Lemons,
D.\,S., Jones, M.\,E., and Mostrom, M.\,A.: 1981, \emph{Phys.
Fluids} {\bf 24}, 1818.
\bibitem[Cattaert et al. (2007)]{cat07}Cattaert, T., Hellberg, M.\,A.,
and Mace, R.\,L.: 2007, \emph{Phys. Plasmas} {\bf 14}, 082111.
\bibitem[Chateau and Meyer-Vernet (1991)]{cha91}Chateau, Y.\,F., and
Meyer-Vernet, N.: 1991, \emph{J. Geophys. Res.} {\bf 96}, 5825.
\bibitem[Chotoo et al. (2000)]{Chot00}Chotoo, K., N. Schwadron, G. Mason, T. Zurbuchen, G. Gloeckler, A. Posner, L. Fisk, A. Galvin, D. Hamilton, and M. Collier: 2000,  J. Geophys. Res., 105(A10), 23107.
\bibitem[Christon (1987)]{Chri87}Christon, S.\,P.: 1987,
\emph{Icarus} {\bf 71}, 448.
\bibitem[Christon et al. (1988)]{Chri88}Christon, S.\,P., Mitchell,
D.\,G., Williams, D.\,J., Frank, L.\,A., Huang, C.\,Y., and Eastman,
T.\,E.: 1988, {\it J. Geophys. Res.} {\bf 93}, 2562.
\bibitem[Christon et al. (1989)]{Chri89}Christon, S.\,P., Williams,
D.\,J., Mitchell, D.\,G., Frank, L.\,A., and Huang, C.\,Y.: 1989,
{\emph J. Geophys. Res.} {\bf 94}, 13409.

\bibitem[Chuang and Hau (2009)]{chu09}Chuang, S.-H., and Hau, L.-N.:
2009, \emph{Phys. Plasmas} {\bf 16}, 022901.
\bibitem[Collier (1993)]{Coll93}Collier, M.\,C.: 1993, {\it Geophys.
Res. Lett.} {\bf 20}, 1531.
\bibitem[Collier (1995)]{Coll95}Collier, M.\,C.: 1995, {\it Geophys.
Res. Lett.} {\bf 22}, 2673.
\bibitem[Collier (2004)]{Coll04}Collier, M.\,C.: 2004, {\it Adv.
Space Res.} {\bf 33}, 2108.
\bibitem[Collier and Hamilton (1995)]{Colal95}Collier, M.\,R., and
Hamilton, D.\,C.: 1995, {\it Geophys. Res. Let.} {\bf 22}, 303.
\bibitem[Collier et al. (1996)]{Coll96}Collier M.\,D., Hamilton,
D.\,C., Gloeckler, G., Bochsler, P., and Sheldon, R.\,B.: 1996,
\emph{Geophys. Res. Lett.} {\bf 23}, 1191.

\bibitem[\protect\citeauthoryear{Dasso \etal}{2003}]{das03}Dasso, S.,
Gratton, F.\,T., and Farugia, C.\,J.: 2003, \emph{J. Geophys. Res.}
{\bf 108}, 1149.
\bibitem[\protect\citeauthoryear{Decker}{1995}]{db95}Decker, D.\,T.,
Basu, B., Jasperse, J.\,R., Strickland, D.\,J., Sharber, J.\,R., and
Winningham, J.\,D.: 1995, \emph{J. Geophys. Res.} {\bf 100}, 21409.

\bibitem[Decker et al. (2005)]{Deck05} Decker, D.\,B., Krimigis S.M., Roelof E.C., Hill M.E., Armstrong T.P., Gloeckler G., Hamilton D.C., Lanzerotti L.J., Science 23, 309, 2020.
\bibitem[De la Haye et al. (2007)]{Haye07} De la Haye, V., Waite Jr.,
J.\,H., Johnson, R.\,E., Telle R.\,V., Cravens, T.\,E., Luhmann
J.\,G., Kasprzak, W.\,T., Gell, D.\,A., Magee, B., Leblanc, F.,
Michael, M., Jurac, S., and Robertson I.\,P.: 2007, \emph{J.
Geophys. Res.} {\bf 112}, A07309.
\bibitem[Dialynas et al. (2009)]{Dial09}Dialynas K., Krimigis S.M., Mitchemm D.G., Hamilton D.C., Krupp N., and Brandt P.C.: 2009,  \emph{J. Geophys. Res.}
{\bf 114}, A01212.
J.\,D.: 1999, {\it Geophys. Res. Lett.} {\bf 23}, 3537.
\bibitem[Dorelli and Scudder (1999)]{Dore99}Dorelli J.\,C., and Scudder,
J.\,D.: 1999, {\it Geophys. Res. Lett.} {\bf 23}, 3537.
\bibitem[Esser and Edgar (2000)]{Esse00}Esser R., and Edgar R.\,J.: 2000, {\it ApJ. (Letters)} {\bf 532}, L71.
\bibitem[Fainberg et al. (1996)]{fain96}Fainberg, J., Osherovich, V.\,A.,
Stone, R.\,G., MacDowall, R.\,J., and Balogh, A.: 1996, in D.
Winterhalter, J.\,T. Gosling, S.\,R. Habbal, W.\,S. Kurth and M.\,N.
Eugebauer, (eds.), {\it Proc. Solar Wind 8 Conference}, AIP
Conference Proceedings {\bf 382}, Melville, NY, US, p. 554.
\bibitem[Feldman et al. (1974)]{fe74}Feldman, W.\,C., Asbridge, J.\,R.,
Bame, S.\,J., and Montgomery, M.\,D.: 1974, {\it Rev. Geophys.} {\bf
12}, 715.
\bibitem[Feldman et al. (1975)]{fe75}Feldman, W.\,C., Asbridge, J.\,R.,
Bame, S.\,J., Montgomery, M.\,D., and Gary, S.\,P.: 1975, {\it J.
Geophys. Res.} {\bf 80}, 4181.
\bibitem[\protect\citeauthoryear{Fisk and Gloeckler}{2006}]{fg06}Fisk,
L.\,A., and Gloeckler, G.: 2006, {\it Astrophys. J.} {\bf 640}, L79.

\bibitem[\protect\citeauthoryear{Fisk and Glockler}{2007}]{fg07}Fisk,
L.\,A., and Gloeckler, G.: 2007, {\it Space Sci. Rev.} {\bf 130}, 153.
\bibitem[Formisano et al. (1973)]{Form73}Formisano, V., Moreno, G.,
Palmiotto, F., Hedgecock, P.\,C.: 1973, {\it J. Geophys. Res.} {\bf
78}, 3714.
\bibitem[\protect\citeauthoryear{Fried and Conte}{1961}]{fc61}Fried, B.\,D.,
and Conte, S.\,D.: 1961, {\it The Plasma Dispersion Function},
Academic Press, New York, US.
\bibitem[Garrett and Hoffman (2000)]{Garr00}Garrett, H.\,B., and Hoffman,
A.\,R.: 2000, {\it IEEE Trans. Plasm. Sci.} {\bf 28}, 2048. M.
\bibitem[Gedalin \etal (2001)]{Ged01}Gedalin, M., Lyubarsky,  Yu.\,E., Balikhin,
M., and Russell, C.\,T.: 2001, {\it Phys. Plasmas} {\bf 8}, 2934.
\bibitem[Gloeckler et al. (1992)]{Gloe92}Gloeckler, G., Geiss, J., Balsiger,
H., Bedini, P., Cain, J.\,C., et al.: 1992, {\it Astron. Astrophys.
Suppl. Ser.} {\bf 92}, 267.
\bibitem[\protect\citeauthoryear{Gloeckler and Fisk}{2006}]{gf06}Gloeckler,
G., and Fisk, L.\,A.: 2006, {\it Astrophys. J.} {\bf 648}, L63.

\bibitem[Gloeckler and Hamilton (1987)]{Gloe87}Gloeckler, G., and Hamilton,
D.\,C.: 1987, {\it Phys. Scripta} {\bf T18}, 73.

\bibitem[Gosling et al. (1993)]{go93}Gosling, J.\,T., Bame, S.\,J.,
Feldman, W.\,C., McComas, D.\,J., and Phillips, J.\,L.: 1993, {\it
Geophys. Res. Lett.} {\bf 20}, 2335.
\bibitem[Hasegawa et al. (1985)]{Hase85}Hasegawa, A., Mima, K., and Duong-van,
M.: 1985, {\it Phys. Rev. Lett.} {\bf 54}, 2608.
\bibitem[Hau and Fu (2007)]{Hau07}Hau, L.\,-N., and Fu, W.\,-Z.: 2007, {\it
Phys. Plasmas} {\bf 14}, 110702.
\bibitem[\protect\citeauthoryear{Hellberg and Mace}{2002}]{hm02}Hellberg,
M.\,A., and Mace, R.\,L.: 2002, {\it Phys. Plasmas} {\bf 9}, 1495.

\bibitem[\protect\citeauthoryear{Hellberg \etal}{2000a}]{hel00}Hellberg,
M.\,A., Mace, R.\,L., Armstrong, R.\,J., and Karlstad, G.: 2000, {\it J. Plasma Phys.} {\bf 64}, 433.

\bibitem[Hellberg et al. (2005)]{hell06} Hellberg M., Mace R., and
Cattaert T.: 2005, {\it Space Sci. Rev.} {\bf 121}, 127.

\bibitem[\protect\citeauthoryear{Hellberg \etal}{2000b}]{hmv00}Hellberg, M.\,A.,
Mace, R.\,L., and Verheest, F.: 2000, in F. Verheest, M. Goosens,
M.\,A. Hellberg, and R. Bharuthram (eds.), {\it Waves in Dusty,
Solar and Space Plasmas, Leuven, 2000}, AIP Conference Proceedings
{\bf 537}, Melville, NY, US, p. 348.
\bibitem[\protect\citeauthoryear{Hellberg et al.}{2009}]{hel09}Hellberg, M.\,A.,
Mace, R.\,L., Baluku, T.\,K., Kourakis, I., and Saini, N.\,S.: 2009,
{\it Phys. Plasmas} {\bf 16}, 094701.
\bibitem[Heerikhuisen et al. (2008)]{heer09} Heerikhuisen, J., Pogorelov, N.\,V.,
Florinski, V., Zank, G.\,P., and Roux, J.\,A.: 2008, {\it Astrophys.
J.} {\bf 682}, 679.
\bibitem[Jung and Hong (2000)]{jung00}Jung, Y.-D., and Hong, W.: 2000,
{\it J. Plasma Phys.} {\bf 63}, 79.
\bibitem[Kamel \etal (2009)]{kam09} Kamel, A., Tribeche, M., and
Zerguini, T.\,H.: 2009, {\emph Phys Plasmas} {\bf 16}, 083701.
\bibitem[Kasparova and Karlicky (2009)]{kasp09}Kasparova, J., and Karlicky, M.: 2009,
{\it Astron. Astrophys.} {\bf 497}, L13.
\bibitem[Kim et al. (2004)]{Kim04}Kim, Y.\,-W., Song, M.\,-Y., and Jung, Y.\,-D.: 2004,
{\it Phys. Scr.} {\bf 70}, 361.
\bibitem[Kletzing (2003)]{Klet03}Kletzing, C.\,A.: 2003, {\it J. Geophys. Res.}
{\bf 108}, 1360.
\bibitem[Ko et al. (1996)]{Koal96}Ko, Y.\,-K., Fisk, L.\,A., Gloeckler, G.,
and Geiss, J.: 1996, {\it Geophys. Res. Let.} {\bf 20}, 2785.
\bibitem[Krimigis et al. (1981)]{Krim81}Krimigis, S.\,M., Carbary, J.\,F.,
Keath, E.\,P., Bostrom, C.\,O., Axford, W.\,I., Gloeckler, G.,
Lanzerotti L.\,J., and Armstrong, T.\,P.: 1981, {\it J. Geophys.
Res.} {\bf 86}, 8227.
\bibitem[Krimigis et al. (1983)]{Krim83}Krimigis, S.\,M., Carbary, J.\,F.,
Keath, E.\,P., Armstrong, T.\,P., Lanzerotti L.\,J., and Gloeckler,
G.: 1983, {\it J. Geophys. Res.} {\bf 88}, 8871.
\bibitem[Krimigis et al. (1986)]{Krim86}Krimigis, S.\,M., Armstrong, T.\,P.,
Axford, W.\,I., Cheng, A.\,F., Gloeckler, G., Hamilton, D.\,C., et
al.: 1986, {\it Science} {\bf 233}, 97.
\bibitem[Lamy et al. (2003a)]{Lamy03}Lamy H., Pierrard, V., Maksimovic, M.,
and Lemaire, J.: 2003a, {\it J. Geophys. Res.} {\bf 108}, 1047.
\bibitem[Lamy et al. (2003b)]{Lamy03b} Lamy H., M. Maksimovic, V. Pierrard, and J. Lemaire: 2003b,  Proceedings of the Tenth International Solar Wind Conference, edited by M. Velli, R. Bruno and F. Malara, American Institute of Physics, 367.
\bibitem[Landi and Pantellini (2001)]{Land01}Landi, S., and Pantellini, F.\,G.: 2001, {\it Astron. Astrophys.} {\bf 372}, 686.
\bibitem[Langmayr et al. (2005)]{lang05}Langmayr, D., Biernat, H.\,K., and
Erkaev, N.\,V.: 2005, {\it Phys. Plasmas} {\bf 12}, 102103.

\bibitem[Lazar and Poedts (2009)]{laza09}Lazar, M., and Poedts S.: 2009, {it
Astron. Astrophys.} {\bf 494}, 311.
\bibitem[Lazar et al. (2008a)]{lspt08}Lazar, M., Schlickeiser, R.,
Poedts, S., and Tautz, R.\,C.: 2008, {\it Mon. Not. R. Astron. Soc.}
{\bf 390}, 168.
\bibitem[Lazar et al. (2009a)]{ltsp09}Lazar, M., Tautz, R.\,C., Schlickeiser,
and R., Poedts, S.: 2009, {\it Mon. Not. R. Astron. Soc.} {\bf 401},
362.
\bibitem[Lazar et al. (2008b)]{lss08}Lazar, M., Schlickeiser, R., and
Shukla, P.\,K.: 2008, {\it Phys. Plasmas} {\bf 15}, 042103.

\bibitem[Le Chat et al. (2009)]{lechat09} Le Chat, G., Issautier, K.,
Meyer-Vernet, N., Zouganelis, I., Maksimovic, M.,
and Moncuquet, M.: 2009, {\it Phys. Plasmas} {\bf 16}, 102903.


\bibitem[Lemaire and Pierrard (2001)]{LePi01}Lemaire, J. and Pierrard, V.: 2001,
{\it Astrophys. Space Sci.} {\bf 277}, 169.
\bibitem[Lee (2007)]{Lee07} Lee, M.\,-J.: 2007, {\it Phys. Plasmas} {\bf 14}, 032112.
\bibitem[\protect\citeauthoryear{Leubner}{1982}]{l82}Leubner, M.\,P.: 1982,
{\it J. Geophys. Res.} {\bf 87}, 6331.
\bibitem[\protect\citeauthoryear{Leubner}{1983}]{l83}Leubner, M.\,P.: 1983, {\it J. Geophys. Res.} {\bf 88}, 469.
\bibitem[\protect\citeauthoryear{Leubner}{2000}]{l00} Leubner, M.\,P.: 2000,
{\it Planet. Space Sci.} {\bf 48}, 133.
\bibitem[\protect\citeauthoryear{Leubner}{2002}]{Leub02} Leubner, M.\,P.: 2002,
{\it Astrophys. Space Sci.} {\bf 282}, 573.
\bibitem[\protect\citeauthoryear{Leubner}{2004a}]{Leub04}Leubner,
M.\,P.: 2004a, {\it Phys. Plasmas} {\bf 11}, 1308.
\bibitem[\protect\citeauthoryear{Leubner}{2004b}]{Leub04b}Leubner, M.\,P.: 2004b,
{\it Astrophys. J.} {\bf 604}, 469.
\bibitem[\protect\citeauthoryear{Leubner}{2008}]{Leub08}Leubner, M.\,P.: 2008,
{\it Nonlin. Proc. Geophys.} {\bf 15}, 531.
\bibitem[\protect\citeauthoryear{Leubner and Schupfer}{2000}]{Leubsc00} Leubner,
M.\,P., and Schupfer, N.: 2000, {\it J. Geophys. Res.} {\bf 105},
27387.
\bibitem[\protect\citeauthoryear{Leubner and Schupfer}{2001}]{Leubsc01} Leubner,
M.\,P., and Schupfer, N.: 2001, {\it J. Geophys. Res.} {\bf 106},
12993.
\bibitem[\protect\citeauthoryear{Leubner and Schupfer}{2002}]{Leubsc02} Leubner,
M.\,P., and Schupfer, N.: 2002, {\it Nonlin. Processes Geophys.}
{\bf 9}, 75.
\bibitem[\protect\citeauthoryear{Leubner and Vi\~nas}{1986}]{lv86} Leubner,
M.\,P., and Vi\~nas, A.\,F.: 1986, {\it J. Geophys. Res.} {\bf 91},
13,366.
\bibitem[\protect\citeauthoryear{Leubner and Voros}{2005}]{Leub05} Leubner, M.\,P.,
and Voros, Z.: 2005, {\it Astrophys. J.} {\bf 618}, 547.
\bibitem[Lie-Svendsen et al. (1997)]{Lies97}Lie-Svendsen, O., Hansteen, V.\,H.,
and Leer, E.: 1997, {\it J. Geophys. Res.} {\bf 102}, 4701.
\bibitem[Livadiotis and McComas (2009)]{Liva09}Livadiotis, G., and McComas, D. J.: 2009, {\it J. Geophys. Res.} {\bf 114}, A11105.
\bibitem[\protect\citeauthoryear{Ma and Summers}{1998}]{ms98} Ma, C., and Summers,
D.: 1998, {\it Geophys. Res. Lett.} {\bf 25}, 4099.
\bibitem[\protect\citeauthoryear{Ma and Summers}{1999}]{Masu99} Ma, C., and Summers,
D.: 1999, {\it Geophys. Res. Lett.} {\bf 26}, 1121.
\bibitem[\protect\citeauthoryear{Mace}{1996}]{mace96}Mace, R.\,L.: 1996,
{\it Phys. Scr.} {\bf T63}, 207.
\bibitem[\protect\citeauthoryear{Mace}{1998}]{m98}Mace, R.\,L.: 1998, {\it J.
Geophys. Res.} {\bf 103}, 14643.
\bibitem[\protect\citeauthoryear{Mace}{2003}]{mace03}Mace, R.\,L.:
2003, {\it Phys. Plasmas} {\bf 10}, 2181.
\bibitem[\protect\citeauthoryear{Mace}{2004}]{mace04}Mace, R.\,L.:
2004, {\it Phys. Plasmas} {\bf 11}, 507.
\bibitem[\protect\citeauthoryear{Mace et al.}{1999}]{mace99}Mace, R.\,L.,
Amery, G., and Hellberg, M.\,A.: 1999, {\it Phys. Plasmas} {\bf 6}, 44.
\bibitem[\protect\citeauthoryear{Mace and Hellberg}{1995}]{mh95}Mace, R.\,L.,
and Hellberg, M.\,A.: 1995, {\it Phys. Plasmas} {\bf 2}, 2098.
\bibitem[\protect\citeauthoryear{Mace and Hellberg}{2009}]{mh09}
Mace, R.\,L., and Hellberg,  M.\,A.: 2009,{\it Phys. Plasmas} {\bf
16}, 072113.
\bibitem[\protect\citeauthoryear{Mace et al.}{1998}]{mht98}Mace, R.\,L.,
Hellberg, M.\,A., and Treumann, R.\,A.: 1998, {\it J. Plasma Phys.}
{\bf 59}, 393.
\bibitem[Magnus and Pierrard (2008)]{MaPi08}Magnus, A.\,P., and Pierrard, V.:
2008, {\it Journal of Computational and Applied Mathematics} {\bf
219}, 431.
\bibitem[Maksimovic et al. (1997a)]{Maks97a}Maksimovic, M.,
Pierrard, V., and Riley, P.: 1997, {\it Geophys. Res. Let.} {\bf
24}, 1151.
\bibitem[Maksimovic et al. (1997b)]{Maks97b}Maksimovic, M., Pierrard, V.,
and Lemaire, J.\,F.: 1997, {\it Astron. Astrophys.} {\bf 324}, 725.
\bibitem[Marsch and Livi (1985)]{ml85}Marsch, E., and Livi, S.: 1985, {\it
Phys. Fluids} {\bf 28}, 1379.
\bibitem[Marsch et al. (1982)]{ma82}Marsch, E., Muhlhauser, K.\,H., Schwenn,
R., Rosenbauer, H., Pilipp, W.\,G., and Neubauer, F.\,M.: 1982, {\it
J. Geophys. Res.} {\bf 87}, 52.
\bibitem[Mauk et al. (1991)]{Mauk91}Mauk, B.\,H., Keath, E.\,P., Kane, M.,
Krimigis, S.\,M., Cheng, A.\,F., Acuna, M.\,H., Armstrong, T.\,P.,
and Ness, N.\,F.: 1991, {\it J. Geophys. Res.} {\bf 96}, 19061.
\bibitem[Mauk et al. (2004)]{Mauk04}Mauk, B.\,H. et al.: 2004, {\it J. Geophys. Res.} {\bf 109}, A09S12.
\bibitem[Meyer-Vernet (1999)]{Meye99}Meyer-Vernet, N.: 1999,
{\it Eur. J. Phys.} {\bf 20}, 167.
\bibitem[Meyer-Vernet (2001)]{Meye01}Meyer-Vernet, N.: 2001, {\it Planet.
Space Sci.} {\bf 49}, 247.
\bibitem[Meyer-Vernet (2007)]{Meye07}Meyer-Vernet, N.: 2007, {\it  Atmosph. Spa. Sc. Series} Cambridge, 463 p.
\bibitem[Meyer-Vernet et al. (1995)]{Meye95}Meyer-Vernet, N., Moncuquet, M.,
and Hoang, S.: 1995, {\it Icarus} {\bf 116}, 202.
\bibitem[\protect\citeauthoryear{Miller}{1991}]{mi91} Miller, J.\,A.,: 1991,
{\it Astrophys. J.} {\bf 376}, 342.
\bibitem[\protect\citeauthoryear{Miller}{1997}]{mi97} Miller, J.\,A.: 1997,
{\it Astrophys. J.} {\bf 491}, 939.
\bibitem[Moncuquet et al. (2002)]{Monc02}Moncuquet, M, Bagenal, F., and
Meyer-Vernet, N.: 2002, {\it J. Geophys. Res.} {\bf 107}, 1260.
\bibitem[Montgomery et al. (1968)]{mo68} Montgomery, M.\,D., Bame, S.\,J., and
Hundhause, A.\,J.: 1968, {\it Geophys. J. Res.} {\bf 73}, 4999.
\bibitem[Moore and Mendillo (2005)]{Moor05} Moore, L., and Mendillo, M.: 2005,
{\it J. Geophys. Res.} {\bf 110}, A05310.
\bibitem[Nieves-Chinchilla and Vinas (2008)]{Niev08} Nieves-Chinchilla, T.,
and Vi\~nas, A.\,F.: 2008, {\it Geofisica Internacional} {\bf 47},
245.
\bibitem[Olsson and Janhunen (1998)]{Olss98}Olsson, A., and Janhunen, P.: 1998, {\it
Ann. Geoph.} {\bf 16}, 298.
\bibitem[Pandey and Pandey (2008)]{Pand08}Pandey, R.\,S., and Pandey, R.\,P.: 2008,
{\it Prog. in Electromagn. Res. C} {\bf 2}, 217.
\bibitem[Pierrard (1996)]{Pier96}Pierrard V.: 1996, {\it J. Geophys. Res.} {\bf 101}, 2669.
\bibitem[Pierrard (2009)]{Pier09}Pierrard V.: 2009, {\it Planet. Space Sci.} {\bf 57},
1260.
\bibitem[Pierrard et al. (2007)]{PiKa07} Pierrard, V., Khazanov, G.\,V., and Lemaire, J.:
2007, {\it J. Atmosph. Sol. Terr. Phys.} {\bf 69} issue 12, Recent
Advances in the Polar Wind Theories and Observations, 2048-2057,
guest editors: Tam, Pierrard and Schunk.
\bibitem[Pierrard and Lamy (2003)]{Pi03}Pierrard, V., and Lamy, H.: 2003, {\it
Solar Physics} {\bf 216}, 47.
\bibitem[Pierrard et al. (2004)]{PiLa04}Pierrard, V., Lamy, H., and Lemaire, J.: 2004,
{\it J. Geophys. Res.} {\bf 109}, A02118.
\bibitem[Pierrard and Lemaire (1996a)]{PiLe96a}Pierrard, V., and Lemaire, J.: 1996a,{\it J.
Geophys. Res.} {\bf 101}, 7923.
\bibitem[Pierrard and Lemaire (1996b)]{PiLe96b}Pierrard, V., and Lemaire, J.:
1996b, {\it Radiat. Meas.} {\bf 26}, 333.
\bibitem[Pierrard and Lemaire (1998)]{PiLe98}Pierrard, V., and Lemaire, J.: 1998,
{\it J. Geophys. Res.} {\bf 103}, 4117.
\bibitem[Pierrard and Lemaire (2001)]{PiLe01}Pierrard, V., and Lemaire, J.: 2001,
{\it J. Atmospheric and Solar-Terrestrial Physics} {\bf 63}, 1261.
\bibitem[Pierrard et al. (1999)]{pml99}Pierrard, V., Maksimovic, M., and
Lemaire, J.\,F.: 1999, {\it J. Geophys. Res.} {\bf 104}, 17021.
\bibitem[Pierrard et al. (2001a)]{pml01}Pierrard, V., Maksimovic, M., and Lemaire,
J.\,F.: 2001, {\it J. Geophys. Res.} {\bf 106}, 29,305.
\bibitem[Pierrard et al. (2001b)]{Pier01b}Pierrard, V., Maksimovic, M., and Lemaire, J.:
2001, {\it Astrophys. Space Sci.} {\bf 277}, 195.
\bibitem[Pierrard and Stegen (2008)]{PiSt08}Pierrard, V., and Stegen, K.: 2008, {\it J.
Geophys. Res.} {\bf 113}, A10209.
\bibitem[Pilipp et al. (1987)]{pi87} Pilipp, W.\,G., Miggenrieder, H., Montgomery, M.\,D.,
Muhlhauser, K.\,H., Rosenbauer, H., and Schwenn, R.: 1987, {\it J. Geophys. Res.} {\bf 92}, 1075.
\bibitem[Pilipp et al. (1990)]{pi90} Pilipp, W.\,G., Miggenrieder, H., Montgomery, M.\,D.,
Muhlhauser, K.\,H., Rosenbauer, H., and Schwenn, R.: 1990, {\it J. Geophys. Res.} {\bf 95}, 6305.

\bibitem[\protect\citeauthoryear{Podesta}{2005}]{pod05}Podesta, J.\,J.: 2005,
{\it Phys. Plasmas} {\bf 12}, 052101.
\bibitem[\protect\citeauthoryear{Podesta}{2008}]{pod08}Podesta, J.\,J.: 2008,
{\it Phys. Plasmas} {\bf 15}, 122902.

\bibitem[Pokhotelov et al. (2002)]{Pokh02}Pokhotelov, O.\,A., Treumann, R.\,A., Sagdeev, R.\,Z.,
Balikhin, M.\,A., Onishchenko, O.\,G., Pavlenko, V.\,P., and Sandberg, I.: 2002, {\it J. Geophys.
Res. A} {\bf 107}, 1312.

\bibitem[Pokhotelov et al. (2008)]{Pokh08}Pokhotelov, O.\,A., Sagdeev, R.\,Z., Balikhin, M.\,A., Onishchenko,
O.\,G., and Fedun, V.\,N.: 2008, {\it J. Geophys. Res.} {\bf 113},
A04225.

\bibitem[Qureshi et al. (2002)]{qure03} Qureshi, M.\,N.\,S.,
Pallocchia, G., Bruno, R., Cattaneo, M.\,B., Formisano, V., R\`eme, H.,
Bosqued, J.\,M., Dandouras, I., Sauvaud, J.\,A., Kistler, L.\,M., M\"obius, E.,
Klecker, B., Carlson, C.,W., McFadden, J.\,P., Parkjs, G.\,K., McCarthy, M.,
Korth, A., Lundin, R., Bologh, A., and Shah, H.\,A.: 2003, in M. Velli, R. Bruno and
F. Malara (eds.), {\it Solar Wind Ten Proc.}, AIP Conference Proceedings {\bf 679}, p. 489.
\bibitem[Qureshi et al. (2006)]{Qure06} Qureshi, M.\,N.\,S.,
Shi, J.\,K., and Ma, S.\,Z.: 2006, in Plasma Sci. IEEE Conf. Rec., 234.
\bibitem[Ralchenko et al. (2007)]{Ralc07}Ralchenko, Yu., Feldman, U., and Doschek, G.\,A.:
2007, {\it Astrophys. J.} {\bf 659}, 1682.
\bibitem[Retherford et al. (2003)]{Reth03}Retherford, K.\,D., Moos, H.\,W., Strobel, D.\,F.:
2003, {\it J. Geophys. Res.} {\bf 108}, 1333.
\bibitem[Saini et al. (2009)]{saini09}Saini, N.\,S., Kourakis, I., and Hellberg, M.\,A.:
2009, {\it Phys. Plasmas} {\bf 16}, 062903.
\bibitem[Saito et al. (2000)]{saito00} Saito, S., Forme, F.\,R.\,E., Buchert, S.\,C.,
Nozawa, S., and Fujii, R.: 2000, {\it Ann. Geoph.} {\bf 18}, 1216.
\bibitem[Salem et al. (2003)]{sa03} Salem, C., Hubert, D., Lacombe, C., Bale, S.\,D., Mangeney, A., Larson, D.\,E.,
Lin, R.\,P.: 2003, {\it Astrophys. J.} {\bf 585}, 1147.
\bibitem[Schippers et al. (2008)]{Schi08}Schippers P., et al.: 2008, {\it J. Geophys. Res.} {\bf 113}, A07208.
\bibitem[\protect\citeauthoryear{Scudder}{1992a}]{Scud92a}Scudder, J.\,D.: 1992a,
{\it Astrophys. J.} {\bf 398}, 299.
\bibitem[\protect\citeauthoryear{Scudder}{1992b}]{Scud92b}Scudder, J.\,D.: 1992b,
{\it Astrophys. J.} {\bf 398}, 319.
\bibitem[\protect\citeauthoryear{Scudder}{1994}]{Scud94}Scudder, J.\,D.: 1994,
{\it Astrophys. J.} {\bf 427}, 446.
\bibitem[\protect\citeauthoryear{Shizgal}{2007}]{sh07} Shizgal, B.\,D.: 2007,
{\it Astrophys. Space Sci.} {\bf 312}, 227.
\bibitem[\protect\citeauthoryear{Shoub}{1983}]{Shou83} Shoub, E.\,C.: 1983,
{\it Astrophys. J.} {\bf 266}, 339.
\bibitem[Singhal and Tripathi (2007)]{sing07}Singhal, R.\,P., and Tripathi, A.\,K.: 2007,
{\it J. Plasma Phys.} {\bf 73}, 207.
\bibitem[Steffl et al. (2004)]{Stef04} Steffl, A.\,J., Bagenal, F., and Stewart, A.\,I.\,F.: 2004,
{\it Icarus} {\bf 172}, 91.
\bibitem[Steinberg et al. (2005)]{Stei05} Steinberg, J.\,T., Gosling, J.T., Skoug R.M., and Wiens, R.\,C.\,F.: 2005,
{\it J. Geophys. Res.} {\bf 110}, A06103.
\bibitem[Stverak et al. (2009)]{Stve09}Stverak, S., Maksimovic M., Travnicek, P.\,M.,
Marsch, E., Fazakerley, A.\,N., and Scime, E.\,E.: 2009, {\it J. Geophys. Res.} {\bf 114}, A05104.
\bibitem[Stverak et al. (2008)]{st08} Stverak, S., Travnicek, P., Maksimovic, M.,
Marsch, E., Fazakerley, A.\,N., and Scime, E.\,E.: 2008, {\it J. Geophys. Res.} {\bf 113}, A03103.

\bibitem[\protect\citeauthoryear{Summers and Ma}{2000}]{sm00} Summers, D., and Ma, C.: 2000, {\it J. Geophys. Res.} {\bf 105}, 15,887.
\bibitem[\protect\citeauthoryear{Summers and Thorne}{1991}]{st91}Summers, D., and Thorne, R.\,M.: 1991, {\it Phys. Fluids B} {\bf 3}, 1835.
\bibitem[\protect\citeauthoryear{Summers and Thorne}{1992}]{st92}Summers, D., and Thorne, R.\,M.: 1992, {\it J. Geophys. Res.} {\bf 97}, 16827.
\bibitem[\protect\citeauthoryear{Summers et al.}{1994}]{st94}Summers, D., Xue, S., and Thorne, R.\,M.: 1994,
{\it Phys. Plasmas} {\bf 1}, 2012
\bibitem[Tam et al. (2007)]{Tam07}Tam, S.\,W.\,Y., Chang, T., and Pierrard, V.: 2007,
{\it J. Atmosph. Sol. Terr. Phys.} {\bf 69}, issue 12, Recent Advances in the Polar Wind
Theories and Observations, p. 1984, guest editors: Tam, Pierrard and Schunk.
\bibitem[\protect\citeauthoryear{Thorne and Summers}{1986}]{ts86}Thorne, R.\,M., and Summers, D.: 1986,
{\it Phys. Fluids} {\bf 29}, 4091.
\bibitem[\protect\citeauthoryear{Thorne and Summers}{1991}]{ts91}Thorne, R.\,M., and Summers, D.: 1991,
{\it Phys. Fluids B} {\bf 3}, 2117.
\bibitem[\protect\citeauthoryear{Treumann}{1997}]{t97}Treumann, R.\,A.: 1997, {\it Geophys. Res. Lett.} {\bf 24},
1727.
\bibitem[\protect\citeauthoryear{Treumann}{1999a}]{treu99a}Treumann, R.\,A.: 1999,
{\it Phys. Scripta} {\bf 59}, 19.
\bibitem[\protect\citeauthoryear{Treumann}{1999b}]{treu99b}Treumann, R.\,A.: 1999,
{\it Phys. Scripta} {\bf 59}, 204.
\bibitem[\protect\citeauthoryear{Treumann}{2001}]{Treu01}Treumann, R.\,A.: 2001,
{\it Astrophys. Space Sci.} {\bf 277}, 81.
\bibitem[\protect\citeauthoryear{Treumann and Jaroschek}{2008}]{treu08}Treumann,
R.\,A., and Jaroschek, C.\,H.: 2008, {\it Phys. Rev. Lett.} {\bf
100}, 155005.
\bibitem[\protect\citeauthoryear{Treumann et al.}{2004}]{treu04}Treumann, R.\,A.,
Jaroschek, C.\,H., and Scholer, M.: 2004, {\it Phys. Plasmas} {\bf 11}, 1317.
\bibitem[\protect\citeauthoryear{Tripathi and Misra}{2000}]{trip00}Tripathi, A.\,K.,
and Misra, K.\,D.: 2000, {\it Earth, Moon and Planets} {\bf 88}, 131.
\bibitem[\protect\citeauthoryear{Tripathi and Singhal}{2008}]{trip08}Tripathi, A.\,K.,
and Singhal, R.\,P.: 2008, {\it Planet. Space Sci.} {\bf 56}, 310.
\bibitem[\protect\citeauthoryear{Tsallis}{1995}]{Tsal95} Tsallis C.: 1995,
{\it Phys. A} {\bf 221}, 277.
\bibitem[\protect\citeauthoryear{Valentini and D'Agosta}{2007}]{vd07}
Valentini, F., and D'Agosta, R.: 2007, {\it Phys. Plasmas} {\bf 14}, 092111.
\bibitem[\protect\citeauthoryear{Vasyliunas}{1968}]{v68}Vasyliunas, V. M.: 1968,
{\it J. Geophys. Res.} {\bf 73}, 2839.
\bibitem[\protect\citeauthoryear{Vi\~nas \etal}{2000}]{Vina00}Vi\~nas, A.\,F., Wong, H.\,K., and Klimas, A.\,J.: 2000,  {\it ApJ.} {\bf 528}, 509.
\bibitem[\protect\citeauthoryear{Vi\~nas \etal}{2005}]{vm05}Vi\~nas, A.\,F., Mace, R.\,L.,
and Benson, R.\,F.: 2005,  {\it J. Geophys. Res.} {\bf 110}, A06202.
\bibitem[Vocks and Mann (2003)]{Vock03}Vocks, C., and Mann, G.: 2003, {\it Astroph. J.} {\bf 593}, 1134.

\bibitem[Vocks et al. (2008)]{Vock08}Vocks, C., Mann, G., and Rausche, G.: 2008, {\it Astron. Astrophys.} {\bf 480}, 527.

\bibitem[Vocks et al. (2005)]{Vock05}Vocks, C., Salem C., Lin R.P., and Mann, G.: 2005, {\it Astroph. J.} {\bf 627}, 540.
\bibitem[Wannawichian et al. (2003)]{wann03}Wannawichian, S., Ruffolo, D., and Kartavykh, Yu.\,Yu.:
2003, {\it Astrophys. J. Suppl. S.} {\bf 146}, 443.
\bibitem[Xiao et al. (1998)]{xts98}Xiao, F., Thorne, R.\,M., and Summers, D.: 1998, {\it Phys. Plasmas} {\bf 5}, 2489.
\bibitem[Xiao (2006)]{xiao06}Xiao, F.: 2006, {\it Plasma Phys. Control. Fusion} {\bf 48}, 203.
\bibitem[Xiao et al. (2006a)]{xial06}Xiao, F., Zhou, Q., He, H., and Tang, L.: 2006, {\it Plasma Phys. Control. Fusion} {\bf 48}, 1437.
\bibitem[Xiao et al. (2006b)]{xiaojgr}Xiao, F., Zhou, Q., Zheng, H., and Wang, S.: 2006,
{\it J. Geophys. Res.} {\bf 111}, A08208.
\bibitem[\protect\citeauthoryear{Xiao et al.}{2007a}]{xiao07}Xiao, F., Zhou, Q., He, H., Tang, L., and Jiayuan, F.: 2007a,
{\it Plasma Sci. Technol.} {\bf 9}, 545.
\bibitem[Xiao et al. (2007b)]{xiao07b}Xiao, F., Zhou, Q., He, H., Zheng, H., and Wang, S.: 2007b, {\it J. Geophys. Res.} {\bf 112}, A07219.
\bibitem[Xiao et al. (2008a)]{xiaoal08}Xiao, F., Chen, L., and Li, J.: 2008, {\it Plasma Phys. Contr. Fusion} {\bf 50}, 105002.
\bibitem[Xiao et al. (2008b)]{xiaoal08b}Xiao, F., Zhou, Q., Li, C., and Cai, A.: 2008,
{\it Plasma Phys Control Fusion} {\bf 50}, 062001.
\bibitem[Xiao et al. (2008c)]{xiao08}Xiao, F., Shen, C., Wang, Y., Zheng, H., and Wang, S.: 2008,
{\it J. Geophys. Res.} {\bf  113}, A05203.
\bibitem[Xue et al. (1993)]{xue93}Xue, s., Thorne, R.\,M., and Summers, D.: 1993,
{\it J. Geophys. Res.} {\bf 98}, 17475.
\bibitem[Xue et al. (1996a)]{xue96a}Xue, s., Thorne, R.\,M., and Summers, D.: 1996a,
{\it J. Geophys. Res.} {\bf 101}, 15457.
\bibitem[Xue et al. (1996b)]{xue96b}Xue, s., Thorne, R.\,M., and Summers, D.: 1996b,
{\it J. Geophys. Res.} {\bf 101}, 15467.

\bibitem[Younsi and Tribeche (2008)]{you08}Younsi, S., and Tribeche, M.: 2008,
{\emph Phys. Plasmas} {\bf 15}, 073706.

\bibitem[\protect\citeauthoryear{Zaheer and Murtaza}{2007}]{zm07}Zaheer, S., and Murtaza, G.: 2007,
{\it Phys. Plasmas} {\bf 14}, 022108.
\bibitem[\protect\citeauthoryear{Zaheer \etal}{2004}]{zah04}Zaheer, S., Murtaza, G., and Shah, H.\,A.: 2004,
{\it Phys. Plasmas} {\bf 11}, 2246.
\bibitem[Zhou et al. (2009)]{zhou09} Zhou, Q.\,H., Jiang, B., Shi, X.\,-H., and Li, J.\,-Q.: 2009,
{\it Chin. Phys. Lett.} {\bf 26} 025201.
\bibitem[Zouganelis (2008)]{zoug08}Zouganelis, I.: 2008, {\it J. Geophys. Res.} {\bf 113}, A08111.
\bibitem[Zouganelis et al. (2004)]{zoug04} Zouganelis, I., Maksimovic, M., Meyer-Vernet, N., Lamy, H.,
and Issautier, K.: 2004, {\it Astrophys. J.} {\bf 606}, 542.
\bibitem[Zouganelis et al. (2005)]{zo05} Zouganelis, I., Meyer-Vernet, N., Landi, S., Maksimovic, M.,
and Pantellini, F.: 2005, {\it Astrophys. J. Lett.} {\bf 626}, L117.

\end{thebibliography}
\end{document}